\documentclass[12pt,preprint]{aastex}
\usepackage{natbib,psfig,emulateapj5}

\newcommand\nhgal{\hbox{{$N_{\rm H}^{\rm Gal}$}}}

\newcommand\msun{\hbox{{$M_{\sun}$}}}

\newcommand\einstein{{\sl Einstein}}
\newcommand\rosat{{\sl ROSAT}}
\newcommand\asca{{\sl ASCA}}

\newcommand\chandra{{\sl Chandra}}
\newcommand\xmm{{\sl XMM}}

\newcommand\xspec{{\sc xspec}}

\newcommand\apec{{\sc apec}}
\newcommand\ciao{{\sc ciao}}
\newcommand\dmfilth{{\sc dmfilth}}
\newcommand\wavdetect{{\sc wavdetect}}

\newcommand\kmsmpc{{\rm km s$^{-1}$ Mpc$^{-1}$}}
\newcommand\cmsq{{\rm cm$^{-2}$}}

\newcommand\lb{\hbox{{$L_{\rm B}$}}}
\newcommand\lsun{\hbox{{$L_{\sun}$}}}

\newcommand\solar{\hbox{{$Z_{\odot}$}}}

\newcommand\ex{\hbox{{$\epsilon_{\rm x}$}}}
\newcommand\pa{\hbox{{$\rm PA$}}}
\newcommand\radpro{\hbox{{$\Sigma_{\rm x}(r)$}}}
\newcommand\sigx{\hbox{{$\Sigma_{\rm x}$}}}
\newcommand\jx{\hbox{{$j_{\rm x}$}}}
\newcommand\ls{\hbox{{$L_{\star}$}}}
\newcommand\phis{\hbox{{$\Phi_{\star}$}}}
\newcommand\eopt{\hbox{{$\epsilon_{\rm opt}$}}}
\newcommand\es{\hbox{{$\epsilon_{\star}$}}}
\newcommand\as{\hbox{{$a_{\star}$}}}
\newcommand\rc{\hbox{{$r_{\rm c}$}}}

\newcommand\jxs{\hbox{{$j_{\rm x}^{\star}$}}}
\newcommand\radpros{\hbox{{$\Sigma_{\rm x}^{\star}(r)$}}}
\newcommand\sigxs{\hbox{{$\Sigma_{\rm x}^{\star}$}}}
\newcommand\exs{\hbox{{$\epsilon_{\rm x}^{\star}$}}}

\newcommand\ssigx{\hbox{{$\sigma_{\rm x}$}}}
\newcommand\ssigxs{\hbox{{$\sigma_{\rm x}^{\star}$}}}
\newcommand\chisq{\hbox{{$\chi^2$}}}

\begin{document} 

\title{Chandra Evidence for a Flattened, Triaxial Dark Matter Halo \\ in the
Elliptical Galaxy NGC 720}

\author{David A. Buote\altaffilmark{1}, Tesla
E. Jeltema\altaffilmark{2},  Claude R. Canizares\altaffilmark{2}, \&
Gordon P. Garmire\altaffilmark{3}} 
\altaffiltext{1}{Department of Physics and Astronomy, University of California
at Irvine, 4129 Frederick Reines Hall, Irvine, CA 92697-4575}
\altaffiltext{2}{Center for Space Research and Department of Physics,
Massachusetts Institute of Technology 37-241, 77 Massachusetts Avenue,
Cambridge, MA 02139}
\altaffiltext{3}{The Pennsylvania State University, 525 Davey Lab,
University Park, PA 16802}

\slugcomment{Accepted for Publication in The Astrophysical Journal}

\begin{abstract}

We present an analysis of a \chandra\ ACIS-S observation of the
elliptical galaxy NGC 720 to verify the existence of a dark matter
halo and to measure its ellipticity. The ACIS-S3 image reveals over 60
point sources distributed throughout the field, most of which were
undetected and therefore unaccounted for in previous X-ray
studies. For semi-major axes $a\la 150\arcsec$ ($18.2h_{70}^{-1}$~kpc)
the ellipticity of the diffuse X-ray emission is consistent with a
constant value, $\ex\approx 0.15$, which is systematically less than
the values 0.2-0.3 obtained from previous \rosat\ PSPC and HRI
observations because of the unresolved point sources contaminating the
\rosat\ values. The \chandra\ data confirm the magnitude of the $\sim
20\degr$ position angle (PA) twist discovered by \rosat\ over this
region. However, the twist in the \chandra\ data is more gradual and
occurs at smaller $a$ also because of the point sources contaminating
the \rosat\ values. For $a\ga 150\arcsec$ out to $a=185\arcsec$
($22.4h_{70}^{-1}$~kpc), which is near the edge of the S3 CCD, \ex\
and PA diverge from their values at smaller $a$. Possible origins of
this behavior at the largest $a$ are discussed.

Overall the ellipticities and PA twist for $a\la 150\arcsec$ can be
explained by the triaxial mass model of NGC 720 published by
Romanowsky \& Kochanek (which could not produce the abrupt PA twist in
the \rosat\ HRI data). Since the optical image displays no substantial
isophote twisting, the X-ray PA twist requires a massive dark matter
halo if the hot gas is in hydrostatic equilibrium. Furthermore, the
values of \ex\ obtained by \chandra\ are too large to be explained if
the gravitating mass follows the optical light ($M\propto \ls$)
irrespective of the PA twist: The $M\propto \ls$ hypothesis is
inconsistent with the \chandra\ ellipticities at the 96\% confidence
level assuming oblate symmetry and at the 98\% confidence level for
prolate symmetry. Thus, both the PA twist and the ellipticities of the
\chandra\ image imply the existence of dark matter independent of the
temperature profile of the gas. This geometric evidence for dark
matter cannot be explained by alternative gravity theories such as the
Modification of Newtonian Dynamics (MOND).

To constrain the ellipticity of the dark matter halo we considered
both oblate and prolate spheroidal mass models to bracket the full
range of (projected) ellipticities of a triaxial ellipsoid.  The dark
matter density model, $\rho\propto (a_s^2+a^2)^{-1}$, provides the
best fit to the data and gives ellipticities and $1\sigma$ errors of
$\epsilon=0.37\pm 0.03$ for oblate and $\epsilon=0.36\pm 0.02$ for
prolate models. Navarro-Frenk-White (NFW) and Hernquist models give
similar ellipticities for the dark matter. These moderate
ellipticities for the dark halo are inconsistent with both the nearly
spherical halos predicted if the dark matter is self-interacting and
with the highly flattened halos predicted if the dark matter is cold
molecular gas. These ellipticities may also be too large to be
explained by warm dark matter, but are consistent with galaxy-sized
halos formed in the currently popular $\Lambda$CDM paradigm.

\end{abstract}


\section{Introduction}
\label{intro}

It is now almost 20 years that the ``Cold Dark Matter'' paradigm
(CDM), in which most of the matter in the universe is assumed to be
collisionless dark matter, has been regarded as the standard
cosmological model of the formation of structure in the universe.  The
CDM model (and particularly its variant $\Lambda$CDM) has achieved
many successes in describing current observations, especially with
regard to the cosmic microwave background radiation and the
large-scale clustering properties of galaxies
\citep[e.g.,][]{rees02}. Recently, however, many have drawn attention to
the problems $\Lambda$CDM has describing structure on galaxy scales
\citep[e.g.,][]{sell02a}. 

One problem with CDM is that it predicts that all galaxies, including
elliptical galaxies, should possess massive dark matter halos. But
definitive evidence for such halos in isolated elliptical galaxies has
proven difficult to establish. Stellar dynamical analyses have made
significant progress limiting the uncertainties on the radial mass
profile in Es due to velocity dispersion anisotropy
\citep[e.g.,][]{deze97,gerh02}, but a recent study by
\citet{baes01} argues that interstellar dust in Es renders such
mass determinations unreliable. Gravitational lensing studies have the
potential to provide important constraints on the mass distributions
in elliptical galaxies from statistical analysis of the weak shear
fields of many galaxies \citep[e.g.,][]{nata00}. Statistical averaging
of the properties of strong lenses has not yet provided clear evidence
for dark matter in isolated Es \citep[e.g.,][]{keet98}. 

There is evidence for dark matter in the outer halos ($\ga 100$~kpc)
of elliptical galaxies obtained from kinematic tracers such as
globular clusters, though for the few objects studied the inferred
masses imply a surrounding group or cluster \citep[e.g., in NGC
4472,][]{zepf02}. Accurate constraints on the radial temperature
profiles of hot gas obtained by \rosat\ X-ray observations have
indicated substantial amounts of dark matter for elliptical galaxies
in the centers of groups \citep[e.g., NGC 1399,][]{jone97} and
clusters \citep[e.g., M87,][]{nuls95}. But because they are fainter in
X-rays, \rosat\ (and \einstein, \citealt{fabb89}) did not provide
temperature profiles of sufficient accuracy to clearly demonstrate the
need for dark matter in isolated Es.

It is clear that the evidence for dark matter in isolated elliptical
galaxies just mentioned is not extensive, and even in disk galaxies
the case for dark matter is not considered to be as strong as it once
was \citep[e.g.,][]{evan02}. Even for those galaxies where dark matter
is suggested on the basis of standard Newtonian gravity,
\citet{sell02b} argue that all such evidence for dark matter cannot be
distinguished from general modified gravity theories, such as the
Modification of Newtonian Dynamics (MOND), proposed by
\citet{milg83a,milg83b,milg83c}.

A powerful test for dark matter that can distinguish dark matter from
MOND in isolated Es is the ``Geometric Test for Dark Matter'' we
introduced previously \citep{buot94,buot96a,buot98a}.  The Geometric
Test is a comparison between the ellipticities and orientations of the
observed X-ray isophotes with those expected if the gravitating mass
traces the same shape as the stellar light. Applying this Geometric
Test to the ellipticities of the \rosat\ data of two isolated Es and
one isolated S0 galaxy, we found strong evidence for dark matter for
the E4 galaxy NGC 720 (99\% conf.) and marginal evidence for the S0
galaxy NGC 1332 (90\% conf.) and the E4 galaxy NGC 3923 (80\%-85\%
conf.) -- see \citet{buot98a} for a review. The \rosat\ data of NGC
720 also displayed an X-ray position angle twist. But this twist could
not be explained with triaxial dark matter models
\citep{buot96d,roma98} which cast some doubt on whether the X-ray
emission indeed traces the potential in NGC 720. And since the
evidence for dark matter in NGC 1332 and NGC 3923 is marginal, higher
quality observations of these systems are required to assess more
definitively the existence of dark matter halos.

For the galaxies mentioned above we inferred from the \rosat\
observations that the dark matter ellipticities range from
approximately $0.4-0.6$ with some indication (particularly in NGC 720)
that the mass is flatter than the stellar distribution. These moderate
ellipticities provide important constraints on competing models for
the dark matter, some of which predict spherical halos \citep{sper00},
highly flattened halos \citep{pfen94}, and moderately flattened halos
\citep[e.g.,][]{bull02}. Owing to the small number of measurements of
the flattening of dark halos \citep[e.g.,][]{merr02,sack96}, and that
such previous measurements are of limited accuracy, the different
predictions for halo shapes have not yet been given a definitive
test.

We have obtained a high-resolution \chandra\ observation of NGC 720 to
reexamine the existence of a dark matter halo via the Geometric Test
and to measure its flattening. A major issue to be addressed is the
effect of discrete sources, such as have been found in large
quantities for early-type galaxies with \chandra\
\citep[e.g.,][]{sara00}, on the conclusions drawn from lower
resolution \rosat\ data. Although calculating triaxial models is
beyond the scope of this paper, we will also measure the position
angles of the X-ray isophotes and compare them to the previously
published triaxial model of \citet{roma98}. We assume a distance of
$25$ Mpc to NGC 720 for $H_0=70h_{70}$~\kmsmpc\ (and $\Omega_0=0.3$)
in which case $1\arcsec=0.12$~kpc.

The paper is organized as follows. We discuss the data preparation in
\S \ref{obs}. The treatment of point sources and the measurement of
the X-ray ellipticities, position angles, and radial profile are
discussed in \S \ref{image}. The Geometric Test for dark matter is
performed in \S \ref{geomtest}. We examine the \chandra\ constraints
on the temperature profile in \S \ref{temp} and the ellipticity of the
dark matter in \S \ref{dm}. Finally, in \S \ref{conc} we present our
conclusions.

\section{Observations and Data Reduction}
\label{obs}

NGC 720 was observed with the ACIS-S3 camera for $\approx 40$~ks
during AO-1 as part of the guaranteed-time-observer program (PI
Garmire). The events list was corrected for charge-transfer
inefficiency according to
\citet{town02}, and only events characterized by the standard
\asca\ grades\footnote{http://cxc.harvard.edu/udocs/docs/docs.html}
were used. The standard \ciao\footnote{http://cxc.harvard.edu/ciao/}
software (version 2.2.1) was used for most of the subsequent data
preparation.

Since the diffuse X-ray emission of NGC 720 fills the entire S3 chip,
we used the standard background
templates\footnote{http://cxc.harvard.edu/cal} to model the
background. When attempting to use the {\sc lc\_clean} script to clean
the source events list of flares with the same screening criteria as
the background templates, only 12~ks of data remained. Examination of
the original light curve reveals only one significant flare ($\sim
6$~ks worth of data) near the beginning of the observation, but the
light curve does gently rise and fall during the rest of the
observation so that most of the data does not lie within the standard
3-sigma clip. Consequently, we excluded only the obvious flare leaving
a total exposure of 34.4~ks.

Because the quiescent background can vary typically by $\sim 10\%$
between observations, and we do not use exactly the same screening
criteria as the templates, it is necessary to normalize the
background templates to match the NGC 720 observation. Since NGC 720
has a gas temperature $\sim 0.6$ keV there is little emission from hot
gas for energies $>5$~keV. Consequently, we normalized the S3
background template by comparing source and background counts in the
5-10 keV band extracted from regions near the edges of the S3 field.

We binned the cleaned S3 events lists for the source and background
into images of $1\arcsec\times 1\arcsec$ pixels (actually
$0.984\arcsec\times 0.984\arcsec$ pixels, but we shall round off
throughout the paper). Since background dominates at high energies we
selected events with energies 0.3-3 keV. Likely problems with the
spectral calibration near the low end of the bandpass (see \S
\ref{temp}) are not important for the imaging analysis.

Although vignetting is a small effect within $\approx 3\arcmin$ of the
aim point of the S3 where our analysis is concentrated, we explored
the viability of using an exposure map to flatten the image.  The
inherent difficulty in constructing a single exposure map for the
image of a diffuse source for instruments like ACIS has been discussed
by \citet{davisj01} and Houck\footnote{see
http://cxc.harvard.edu/ciao/documents\_manuals.html}.  Since, however,
the diffuse gas of NGC 720 is remarkably isothermal (see \S
\ref{temp}) we are able to construct a well-defined ``weighted''
exposure map following the procedure on the \ciao\ website. (Note we
find no tangible differences in any results mentioned below if instead
we use a monochromatic exposure map evaluated at 0.77 keV which is the
peak of the counts spectrum.)

Unfortunately, the current method for flat-fielding images introduces
spurious features along node boundaries where the intrinsic exposure
is small relative to the rest of the CCD. When the exposure map is
divided into the raw image to create an exposure-corrected image,
regions of relatively low exposure (e.g., node boundaries, bad
columns, and edges) require large corrections. Any intrinsic errors in
the exposure map in these regions, and any noise in the data, will be
amplified as a result.

For NGC 720 we find that dividing by the exposure map amplifies the
linear distortion along the node 1-2 boundary that reaches within
$\approx 30\arcsec$ of the center of NGC 720. Ellipticities and
position angles computed between radii of $\sim 30\arcsec - 60\arcsec$
are systematically lower by $20\%-30\%$ with respect to their values
obtained from the un-flattened image.  We emphasize that in the raw
image the linear distortion is very faint and very difficult to see,
and imperceptable in the smoothed image displayed in Figure
\ref{fig.image}; i.e., this feature has neglible effect on the
ellipticities and position angles considering the statistical
uncertainties on these quantities (\S \ref{results}).  Consequently,
since we have a case where ``the cure is worse than the disease'',
when computing ellipticities and position angles we do not use the
exposure-corrected image. But since this non-circular, low-surface
brightness feature amplified by the exposure map has no perceptible
effect on the azimuthally averaged radial profile, we do use the
exposure-corrected image to account for the small effects of
(large-scale) telescopic vignetting for analysis of the image radial
profile.

\section{Image Analysis}
\label{image}

\begin{figure*}[t]
\parbox{0.49\textwidth}{
\centerline{\psfig{figure=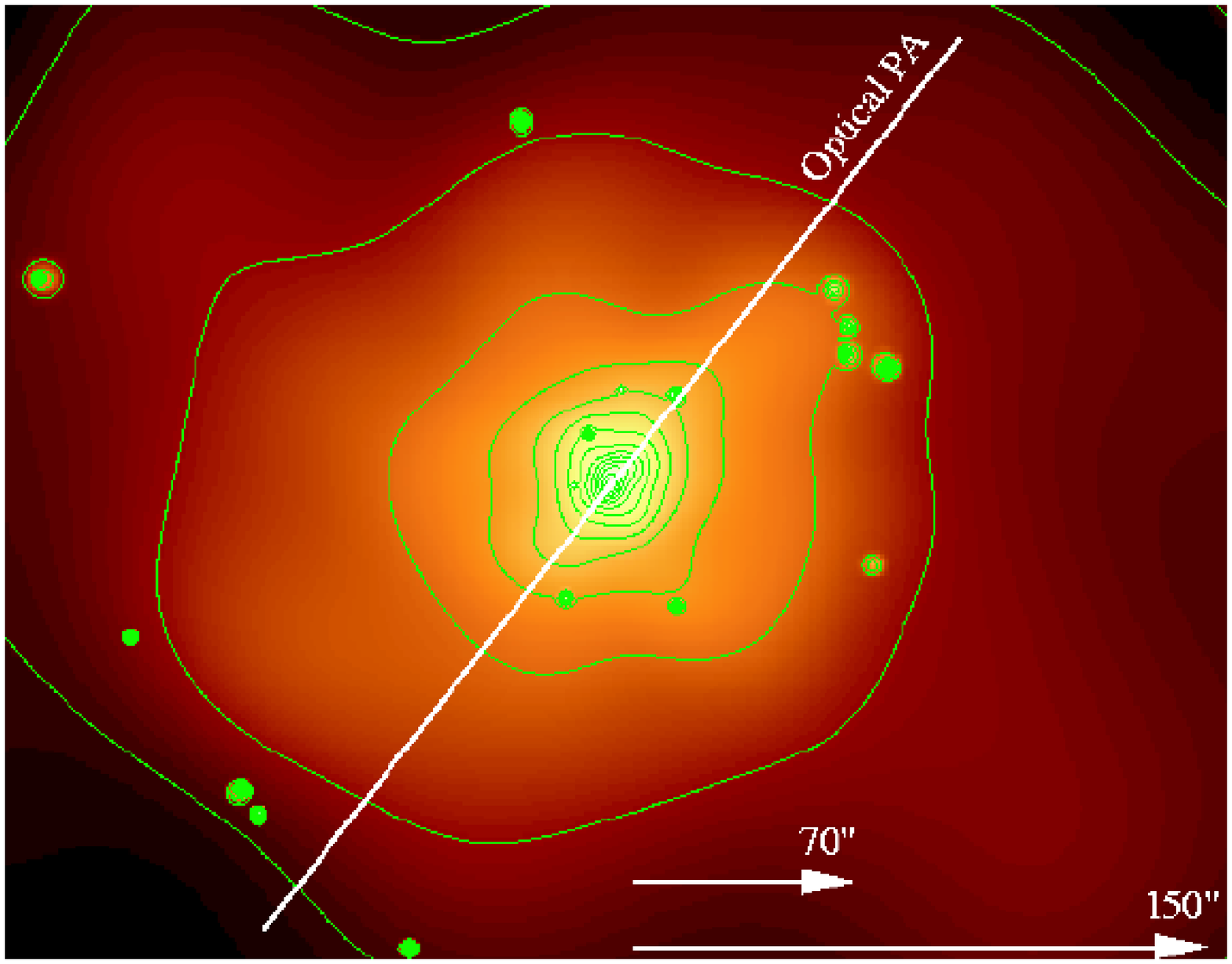,height=0.27\textheight}}
}
\parbox{0.49\textwidth}{
\centerline{\psfig{figure=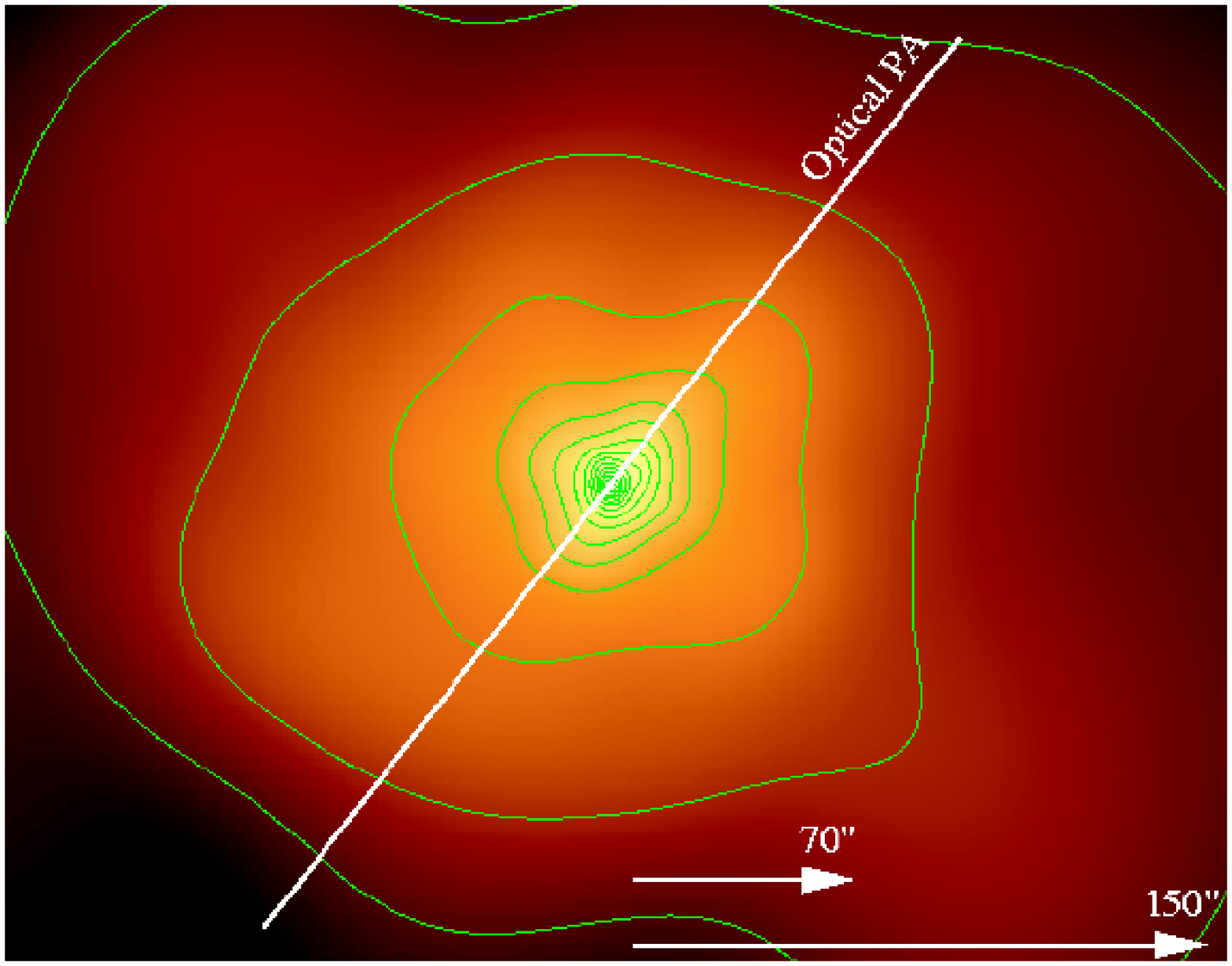,height=0.27\textheight}}
}
\caption{\label{fig.image}  ({\sl Left}) False-color ACIS-S3 image of
NGC 720 in the 0.3-3 keV band. The image has been adaptively smoothed
using the \ciao\ task {\sc csmooth} with default parameters. No
exposure-map correction or background subtraction has been
applied. Contours are spaced according to the square root of the
intensity. ({\sl Right}) Same image except the point sources have been
removed. The contour levels are the same as the original image. Each
image is oriented so that celestial N is up and E is to the left.}
\end{figure*}

In Figure \ref{fig.image} we display the adaptively smoothed S3 image
of NGC 720 within a radius of $150\arcsec$ of the galaxy center with
some isophotal contours overlaid. (This smoothing is done only for
display purposes. All analysis discussed below is performed on the
unsmoothed image; cf. Figure 1 of \citealt{tesla}).  The most striking
difference between the \chandra\ and \rosat\ images
\citep[e.g.,][]{buot98a} is the large number of point sources clearly
detected by \chandra.  Many of these sources are located within
$1\arcmin$ of the galaxy center.

The morphology of the diffuse emission within $\sim 100\arcsec$ of the
galaxy center is similar to that described for previous \rosat\
observations. The isophotes are moderately flattened and the
orientations of the outer isophotes appear to be tilted with respect
to the innermost regions. But caution must be exercised when viewing
this adaptively smoothed image since small-scale features in the
diffuse emission have low statistical significance (especially
apparent wiggles in the isophotes). Below we shall quantify the
morphology of the diffuse emission using moments of the image.

\subsection{Removal of Point Sources}
\label{src}

Before we can analyze the properties of the diffuse emission we must
remove the point sources. (Analysis of the properties of these sources
is presented in a companion paper -- \citealt{tesla}.) However, it is
insufficient only to remove the sources since holes in the image still
affect significantly quantities like the ellipticity that are derived
from second (or higher) moments of the image. To obtain reliable
measurements of the ellipticity and position angle throughout the
source-free image it is necessary to replace each embedded source with
a faithful representation of the local diffuse X-ray emission
surrounding the source.

We located point sources with the \ciao\ task \wavdetect\ (a wavelet
detection algorithm) using the default parameter settings. A total of
64 sources were detected over the entire S3 field excluding the center
of NGC 720. These include the 41 sources within $2.5\arcmin$ of the
galaxy center discussed by \citet{tesla}.

To remove these sources and ``fill'' in the holes with diffuse
emission we used the \ciao\ task \dmfilth. This task takes as input
source and background region definitions. The counts within the source
regions are replaced with values sampled from the background regions.
Within $\sim 3\arcmin$ of the galaxy center we defined the source
extraction regions as circles of radius $2\arcsec$. The corresponding
concentric background regions all have radius $4\arcsec$. The average
background counts in each region was determined by the polynomial
method in \dmfilth. Since in several cases the background region of
one source overlapped another (especially within $\sim 30\arcsec$ of
the galaxy center) we iterated runs of \dmfilth\ using the cleaned
images of previous runs as input to subsequent runs. We found that the
results converged after only three runs.

For radii larger than $\sim 30\arcsec$ we found that \dmfilth\
performed well. In this context ``well'' means that the measured
ellipticity and position angle are smoothly varying functions of
radius; i.e., at the radius of a filled-in source the ellipticity and
position angle do not deviate substantially from values obtained at
adjacent source-free radii. Unfortunately, for radii $\la 30\arcsec$
we found that the ellipticity and position angle did jump
``discontinuously'' near some filled-in sources, especially at the
smallest radii $\la 15\arcsec$ where source background-region overlap
(as noted above) is most pronounced. (Note that such jumps, by
accident, are not apparent for the ellipticity and position angle at
the particular radii $\la 30\arcsec$ plotted below.)

At small radii the relative effect of the sources on derived surface
brightness parameters is necessarily greater because the area of the
source becomes increasingly more significant relative to $r^2$. It is
likely that at such small radii the approach of simply filling in the
source region with a spatially uniform count distribution does not
faithfully (enough) represent the shape the local diffuse emission.
We defer consideration of more sophisticated source-replacement
algorithms to a future study. In our analyses below we treat the
region $r\la 30\arcsec$ with caution.

As should be expected, residual source-replacement effects in these
inner regions are much less important for the radial profile. We shall
therefore use the whole image for construction and analysis of the
radial profile.

\subsection{\ex\ and PA}
\label{ex}

\begin{figure*}[t]
\parbox{0.49\textwidth}{
\centerline{\psfig{figure=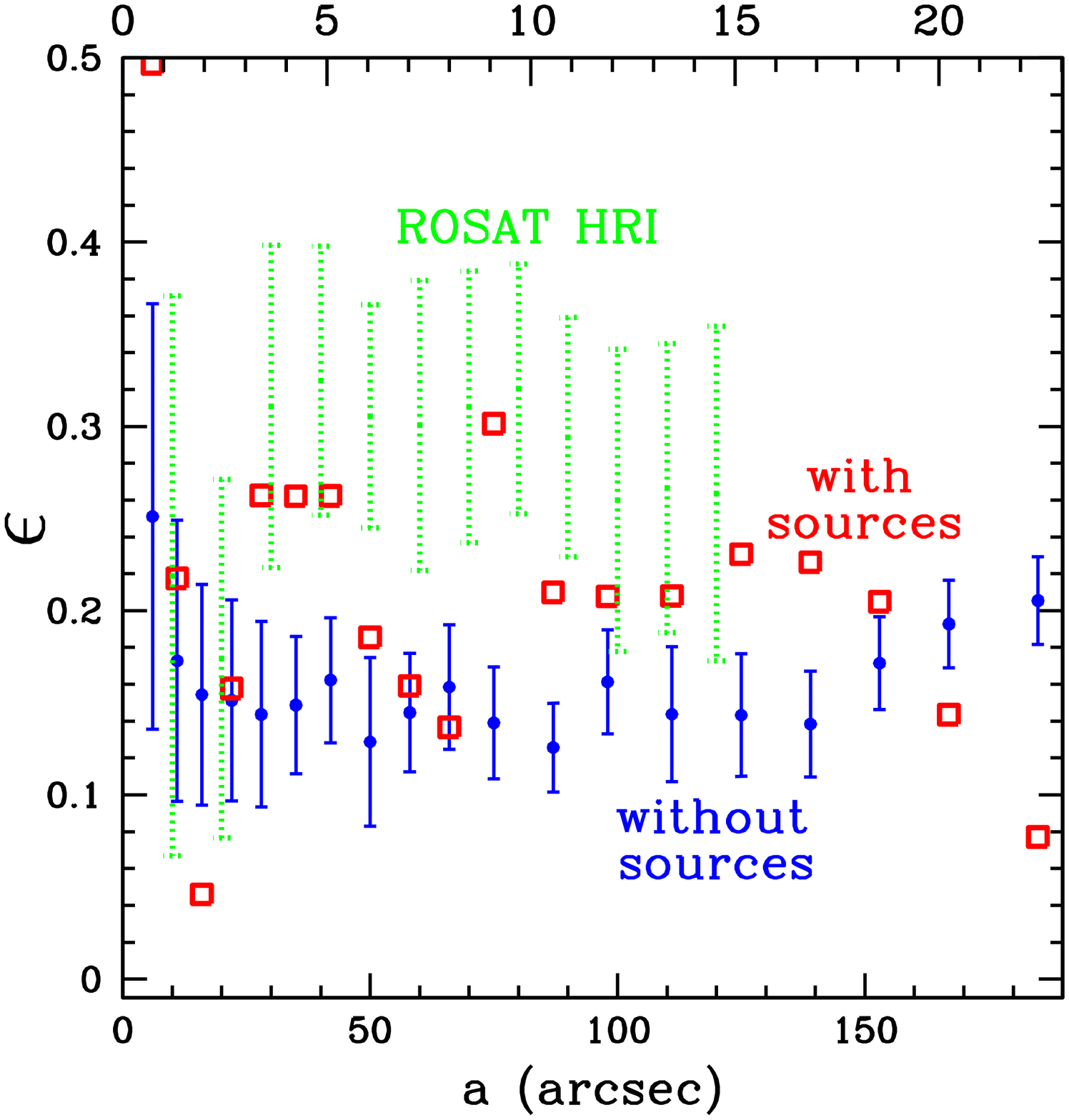,height=0.27\textheight}}
}
\parbox{0.49\textwidth}{
\centerline{\psfig{figure=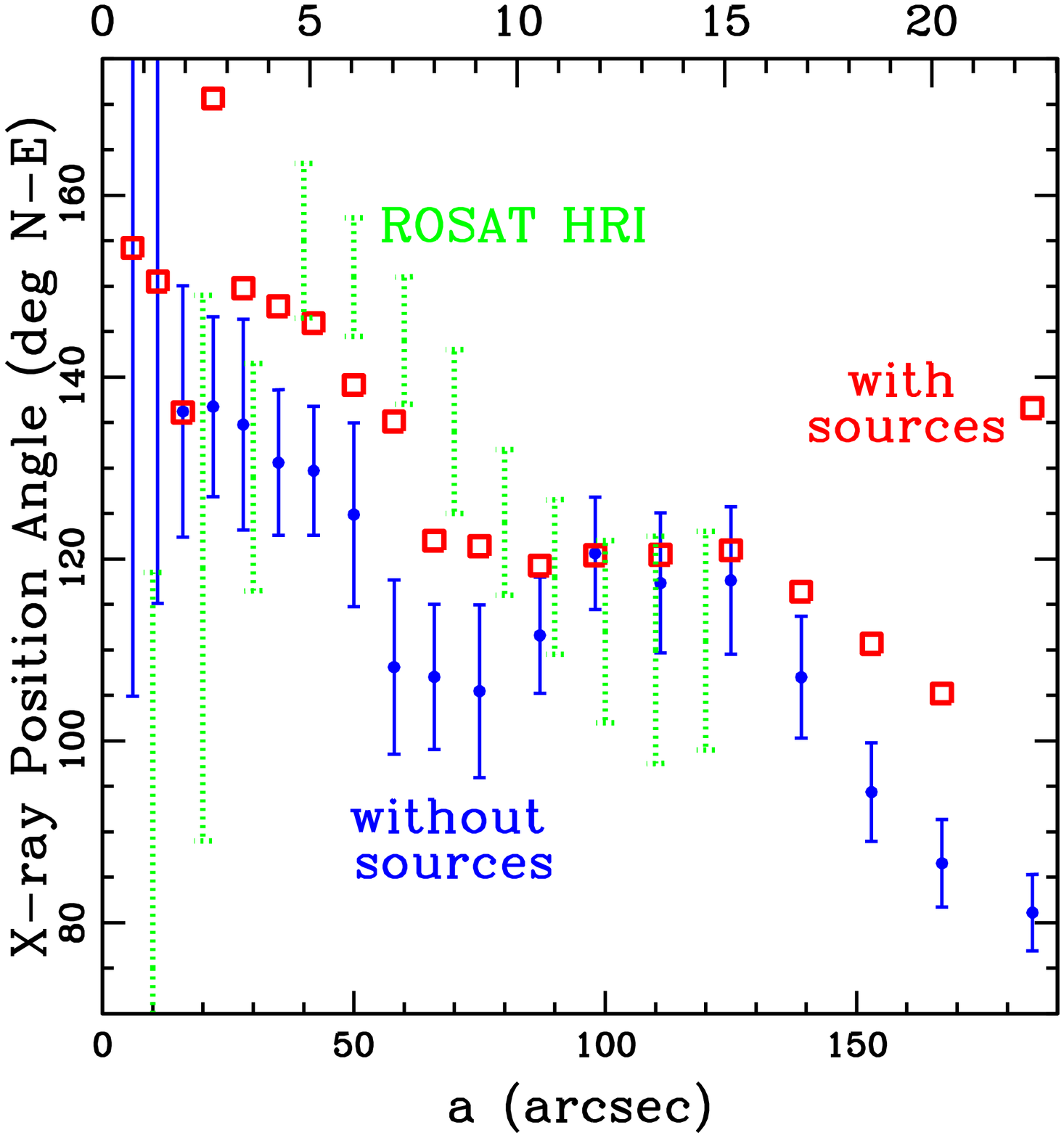,height=0.27\textheight}}
}
\caption{\label{fig.compare} ({\sl Left panel}) \ex\ as a function of
semi-major axis computed from the \chandra\ image with point source
removed (circles and error bars; blue). The (red) boxes are the values
of \ex\ obtained from the image which includes the point sources. The
(green) dotted error bars are the values obtained with the \rosat\ HRI
\citep{buot96d} ({\sl Right panel}) PA as a function of $a$ computed
from the \chandra\ image with point sources removed (circles and error
bars; blue). The (red) boxes are the values of \ex\ obtained from the
image which includes the point sources. Note that because we use all
image pixels interior to $a$ to compute \ex\ and PA, the values and
the error bars for adjacent $a$ are not independent. We express $a$ in
kpc on the top axis.}
\end{figure*}

\begin{figure*}[t]
\parbox{0.49\textwidth}{
\centerline{\psfig{figure=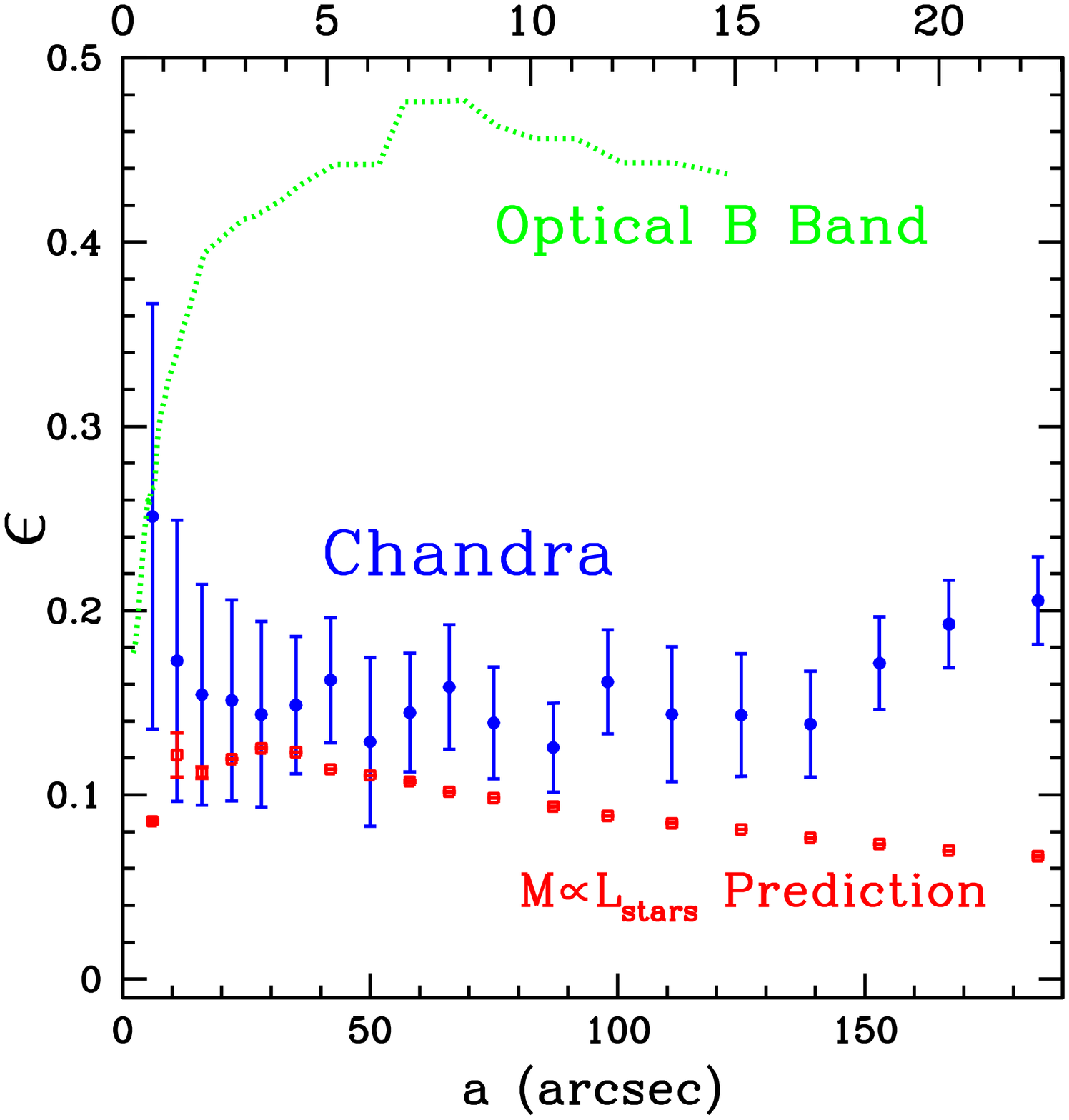,height=0.27\textheight}}
}
\parbox{0.49\textwidth}{
\centerline{\psfig{figure=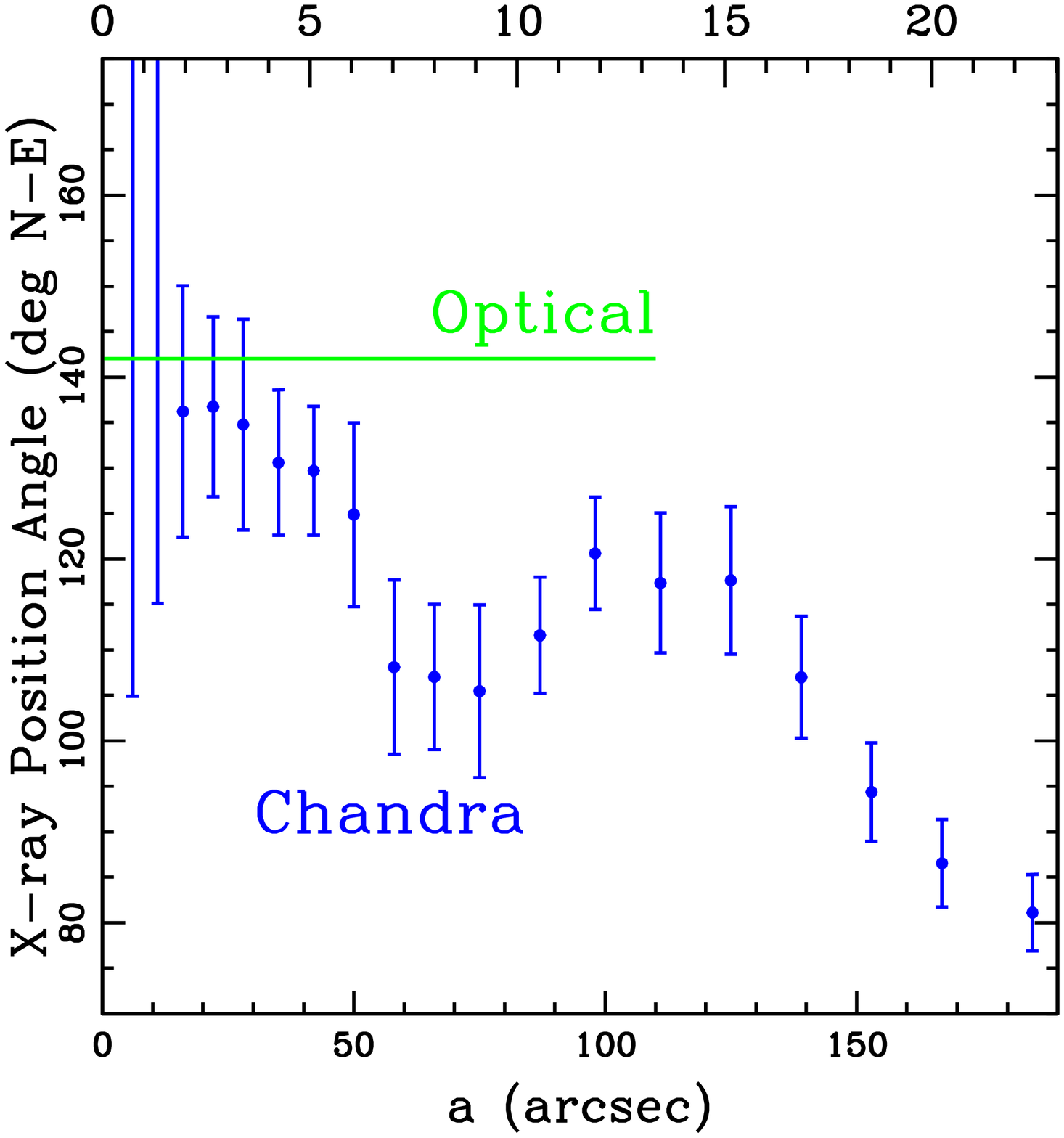,height=0.27\textheight}}
}
\caption{\label{fig.data} ({\sl Left panel}) \ex\ as a function of
semi-major axis computed from the source-free \chandra\ image (circles
and error bars; blue). The optical B-band ellipticity is indicated by
the (green) dashed line (see \S \ref{imp}). The values of \ex\
expected for a model where the gravitating mass follows the optical
light is shown by the (red) squares; i.e., the oblate $M\propto \ls$
model mentioned in the text. ({\sl Right panel}) PA as a function of
$a$ computed from the source-free \chandra\ image (circles and error
bars; blue). The optical B-band value (green) is also shown.  We
express $a$ in kpc on the top axis.}
\end{figure*}

\subsubsection{Method}
\label{method}

The flattening of the X-ray isophotes is of central importance to our
study of the flattening and concentration of the total gravitating
matter in NGC 720. The standard parameter used to denote the
flattening of an individual isophote is the ellipticity, $1 - b/a$,
where $a$ is the semi-major axis and $b$ is the semi-minor
axis. However, fitting perfect ellipses to the X-ray isophotes is not
necessarily justified since the isopotential surfaces generated by an
elliptical mass distribution are not perfect ellipsoids. In addition,
over most of the X-ray image the surface brightness of the diffuse gas
is $\la 1$ counts per pixel. This condition requires that relatively
large areas (in excess of a single-pixel width isophote) must be
averaged over to obtain interesting constraints on the image
flattening.

We quantify the image flattening using the method described by
\citet{cm} and implemented in our previous study of the \rosat\ image
of NGC 720 \citep[e.g. ][]{buot94}. This iterative method is
equivalent to computing the (two-dimensional) principal moments of
inertia within an elliptical region. The ellipticity, \ex, is defined
by the square root of the ratio of the principal moments, and the
position angle, PA, is defined by the orientation of the larger
principal moment.  If these moments are computed within an elliptical
region where the image is perfectly elliptical with constant
ellipticity and orientation, then \ex\ represents a true ellipticity
and PA is the true orientation of the major axis. If the image is not
perfectly elliptical within this region, then \ex\ and PA are average
values weighted heavily toward the edge of the region; i.e., \ex\
provides a useful measure of image flattening which does not assume
the image to be perfectly elliptical.

Following our previous studies we compute $\ex(a)$ using all image
pixels interior to the ellipse defined at $a$ (obtained via
iteration). We did investigate using elliptical annuli to provide a
more direct measurement of the variation of \ex\ with
$a$. Unfortunately, because of the low counts per pixel noted above
the radial fluctuations of \ex, PA, and centroid were considerably
larger when using elliptical annuli.

We estimate uncertainties on \ex\ and PA using a Monte Carlo
procedure. The counts in each pixel are randomized assuming poisson
statistics. Then \ex\ and PA are computed precisely as done for the
original image. After performing 100 such realizations and
corresponding ``measurements'' of the image we compute the standard
deviations of \ex\ and PA at each $a$ for the 100 runs. We take these
standard deviations to be the $1\sigma$ errors.

We have computed \ex\ and PA out to $a=185\arcsec$ from the \chandra\
image such that the background-subtracted counts in each aperture
increase by $\approx 500$ for each $a$. (The edge of the S3 chip
corresponds to $r\approx 215\arcsec$.) Because we use all image pixels
interior to $a$ to compute \ex\ and PA, their values computed for a
given $a$ are correlated with values computed at adjacent inner
$a$. This means that the error bars on \ex\ and PA for adjacent $a$
are not fully independent. But we emphasize that these quantities,
being derived from second moments of the image, are weighted heavily
by the pixels near $a$.

\subsubsection{Results}
\label{results}

In Figure \ref{fig.compare} we show \ex\ and PA computed from the
\chandra\ image before and after removing the point sources. (Note
the error bars for values computed from the image with sources are not
shown for clarity but are of similar magnitude to those obtained from
the source-free image.) For the image with point sources, within $a\la
80\arcsec$ \ex\ varies wildy with $a$ then setlles down to $\ex\approx
0.2$ until it declines for $a\ga 150\arcsec$. This erratic behavior is
a stark constrast to the slowly varying values of \ex\ computed from
the source-free image. {\it We interpret this result as a strong
affirmation of our success at removing and replacing the point sources
over this range in $a$ for the purpose of computing \ex.}

The PA generally does not exhibit such dramatic variations with $a$ in
either case, though the PA values computed from the image with sources
(for the most part) systematically exceed those computed from the
source-free image by $\sim 10\degr$. This relative similarity between
the PAs computed from the raw and source-free images coupled with the
lack of strong fluctuations with $a$ suggest that PA is less sensitive
than is \ex\ to contaminating point sources.  However, with the
present data we cannot rule out the possibility that PA is more
sensitive than \ex\ to contaminating point sources and is therefore
only slightly affected by our attempt to exclude and replace the
sources. Since the values of \ex\ and PA are necessarily interwined,
and we are confident of the success in computing \ex, we do not
believe the PA values computed from the source-free image are still
substantially affected by embedded point sources. Nevertheless, to be
conservative we shall address below the need for dark matter in NGC
720 using \ex\ and PA separately.

Let us focus on the source-free image (Figure \ref{fig.data}).  As
explained above in \S \ref{src} the best-fitting values of \ex\ and PA
computed within $a\la 30\arcsec$ must be considered tentative because
of residual source-replacement errors. Nevertheless, certain trends in
these data appear to be robust. First, for small $a$ we have,
$\ex\approx 0.2-0.3$, consistent with the optical isophotes in this
region within the relatively large estimated errors. Second, the
position angle is consistent with the optical value, $\pa\approx
142\degr$ (N-E), for $a\la 30\arcsec$.

For $30\arcsec\la a\la 150\arcsec$ the ellipticity is consistent with
a constant value, $\ex\approx 0.15$. This value is notably smaller
than the best-fitting values of $\approx 0.20-0.30$ obtained with the
\rosat\ PSPC and HRI \citep{buot94,buot96d,buot98a} between
$30\arcsec\la a\la 110\arcsec$; the HRI values are reproduced in
Figure \ref{fig.compare}. The overall consistency of the \ex\ values
obtained from the \rosat\ image and \chandra\ image with sources
demonstrates that the embedded point sources are responsible for the
\rosat\ values. The most important are the group of point sources
located $\approx 70\arcsec$ to the NW of the galaxy center. These
sources were not identified and removed from the \rosat\ images (and
evaded detection from our symmetry tests because they are located near
the X-ray major axis). Note that the source-free \chandra\ \ex\ values
are smaller than obtained from the HRI for $50\arcsec\la a\la
70\arcsec$. This discrepancy is not resolved by simply degrading the
\chandra\ data by smoothing it with the HRI PSF.  We have not
definitively identified the origin of this discrepancy, but we suspect
it is assocated with the different flux sensitivities of the HRI and
ACIS; i.e., flat spectrum point sources will be more prominent with
respect to the diffuse hot gas in the ACIS data.

The magnitude of the PA twist discovered by \rosat\ is confirmed by
the source-free \chandra\ data, but the precise radial variation of PA
is different because of the unresolved point sources present in the
\rosat\ data. For $80\arcsec\la a\la 150\arcsec$ the position angles,
$\pa\approx 110\degr$, measured by \chandra\ are consistent with the
\rosat\ values. However, for $a\sim 60\arcsec$ the
\rosat\ HRI PA jumps to a value $\pa\approx 140\degr$
whereas the source-free \chandra\ PA increases more gradually from
$\pa\approx 120\degr$ to $\pa\approx 135\degr$ for decreasing $a$
within $a\la 60\arcsec$. The similarity of the PA values obtained for
the \chandra\ image with sources and the HRI demonstrate that the
embedded point sources are responsible for the inflated PA values. The
key point sources for these inflated PA values are located $\sim
45\arcsec$ to the SW of the galaxy center along the minor axis (see
Figure \ref{fig.image} above and Figure 1 of Jeltema et al. 2002).

At large radius ($150\arcsec\la a\la 180\arcsec$) \ex\ increases
steadily to a value of $\approx 0.20$ while the position angle
decreases to a value of $\approx 80\degr$. These changes are highly
significant statistically and are obvious upon inspection of the
smoothed image displayed in Figure \ref{fig.image}. For $a\ga
120\arcsec$ the PSPC and \chandra\ values of \ex\ agree within
errors. This suggests that the effect of unresolved sources on \ex\ is
less important at large radius as might be expected since the PSF FWHM
compared to $a$ decreases with increasing $a$.  However, the PA
measured by the PSPC at large radius does not appear to twist so much
as seen in the \chandra\ data: PSPC PA values are $102\degr\pm
9\degr$ at $a=150\arcsec$ and $107\degr\pm 5\degr$ at
$a=225\arcsec$ \citep{buot94}.

The origin of this \chandra-\rosat\ PA discrepancy at large radius is
unclear, but we would tend to believe the \chandra\ values because of
the issue of point sources in the \rosat\ data. It must be
acknowledged, however, that since the edge of the S3 chip is at
$a\approx 215\arcsec$ there could be problems with flat-fielding which
we think unlikely based on our own investigations with exposure maps
in this region.  It is probably necessary to use the wide-field \xmm\
EPIC CCDs to resolve this issue.

\subsection{Radial Profile}
\label{radpro}

The radial profile ($\radpro$) is the final property derived from
the X-ray image that is required to constrain the shape of the
gravitating matter distribution. We define the image radial profile by
first summing the background-subtracted counts in concentric circular
annuli with radii determined by the semi-major axes used to compute
\ex\ and PA (\S \ref{results}). Then dividing the counts in each
annulus by the area of the annulus we arrive at the radial
profile. The center of each annulus is the same and is taken to be the
centroid computed within a circle of radius $20\arcsec$ placed
initially at X-ray peak. (The centroid varies by $\la 4\arcsec$ for
$r\le 185\arcsec$.) There are $\approx 12000$ background-subtracted
counts within $r=185\arcsec$.

We plot $\radpro$ in Figure \ref{fig.mod}. The radial profile is
quite smooth over the entire range investigated. The somewhat bumpy
HRI profile \citep{buot96d} within $r\la 30\arcsec$ can be attributed
to noise and to the point sources detected in that region by
\chandra. Fitting the \chandra\ profile with a $\beta$ model yields
$r_c\approx 4.6\arcsec$ and $\beta\approx 0.42$ in agreement with the
HRI values. The $\beta$ model is an excellent qualitative (i.e.,
visual) fit to $\radpro$, but because the error bars are very small
the formal quality of the fit is rather poor: $\chi^2=44.3$ for 16
degrees of freedom (dof). The most important fit residuals occur in
the inner bins. These residuals are small when expressed as a ratio of
data to model (i.e., $<10\%$), but are relatively large when expressed
in terms of a $\chi^2$ difference between data model. If, for example,
the inner 3 bins are excluded ($r\la 16\arcsec$) we obtain
$\chi^2=24.9$ for 13 dof.

\section{Geometric Test for Dark Matter}
\label{geomtest}

\subsection{Basic Idea and Assumptions}
\label{idea}

Previously, we have shown that it is possible to test the hypothesis
that the gravitating mass traces the stellar light using only the
X-ray imaging data of the diffuse hot gas -- no knowledge of the value
of the gas temperature or a possible spatial temperature gradient is
required \citep{buot94,buot96a,buot98a}. The key assumptions are that
the hot gas is single-phase and in hydrostatic equilibrium. If these
conditions hold, then the gas volume emissivity, \jx, and the
potential of the gravitating matter, $\Phi$, must have identical
shapes in three-dimensional space (``X-Ray Shape Theorem''); i.e.,
{\it in 3D the surfaces of constant \jx\ are surfaces of constant
$\Phi$ independent of the temperature profile of the gas.}  Therefore,
one may test the hypothesis that the 3D gravitating mass is
distributed in the same way as the 3D stellar light, \ls, obtained
from deprojecting the optical image. This ``Geometric Test for Dark
Matter'' compares the shapes of the isopotential surfaces computed
assuming $M\propto \ls$ to the shapes of the isoemissivity surfaces of
\jx, obtained from deprojecting the X-ray image.

It is important to remember that not only are the shapes of the
isopotential (and therefore isoemissivity) surfaces sensitive to
flattening of the gravitating mass, but they also reflect its radial
distribution. This is illustrated by considering a multipole expansion
of the potential for an arbitrary mass distribution of finite
extent. Near the center of the mass many high-order multipole terms
may contribute resulting in a highly flattened potential in that
region. As one moves away from the center the higher order terms must
decay and eventually (and usually very rapidly) give way to the
spherical monopole term. Hence, an ellipticity gradient in the
potential, and therefore \jx, also probes the radial distribution of
gravitating mass.

This Geometric Test for dark matter that is distributed differently
from \ls\ is also a powerful test of alternative gravity theories such
as MOND.  The fundamental MOND equation is $\mu \vec{g}_{\rm M} =
\vec{g}_{\rm N}$, where $\vec{g}_{\rm N}$ is the Newtonian force
vector, $\vec{g}_{\rm M}$ is the MOND force vector, and $\mu$ is a
scalar function depending only on the ratio $g_{\rm M}/a_0$ -- the
magnitude of the MOND force vector over a constant. Consequently, for
a given mass distribution the shapes of isopotential surfaces in MOND
are the same as the shapes of isopotential surfaces in Newtonian
theory. So if the Geometric Test implies dark matter using Newtonian
theory it follows that dark matter is also required using the MOND
theory. See \citet{buot94} for a more extensive discussion of the
Geometric Test, MOND, and caveats.

The assumption of a single-phase gas for NGC 720 appears to be well
justified. First, the \chandra\ spectrum (see \S \ref{temp}) is
consistent with an isothermal plasma. Second, recent observations with
\chandra\ and \xmm\ have found that even the gas in cooling flow groups and
clusters, systems that might have been expected to be highly
multiphase, is consistent with a single-phase medium \citep[e.g.,
][]{bohr01a,buot02a}.

There are also several lines of evidence which suggest that the
approximation of hydrostatic equilibrium is a good one (see also \S
\ref{he}). Although it is natural to wonder whether the gas is
disturbed by tidal or ram pressure effects, it must be remembered that
NGC 720 is quite isolated from other large galaxies
\citep{dres86}. Aside from the PA twist there are no indications of external
influences in the diffuse X-ray emission; e.g., centroids computed for
the elliptical apertures in \S \ref{results} are all consistent within
$3\arcsec - 4\arcsec$.

We should also emphasize that the wiggles and other irregular features
displayed by the smoothed X-ray isophotes in Figure \ref{fig.image}
are not statistically significant and thus do not necessarily imply
departures of the gas from hydrostatic equilibrium.  However, even if
those wiggles do in fact arise from real non-equilibrium motions in
the gas, the hydrostatic assumption is still very likely a good
one. Previously, in \citet{buot95a} we analyzed the X-ray isophotes of
a galaxy cluster formed in an N-body / hydrodynamical simulation. It
was found that relatively soon after major mergers (i.e., 1-2 crossing
times) the cluster settled down to a quasi-relaxed state where the hot
gas traced approximately the shape of the gravitational potential even
though the shapes of the X-ray isophotes displayed prominent irregular
features.

As in our previous studies we shall assume negligible rotation of the
hot gas. We expect this to be a good approximation for several
reasons. Most importantly, like most massive elliptical galaxies there
is negligible stellar rotation; NGC 720 has a mass-weighted stellar
rotational velocity of 35 km s$^{-1}$ \citep{busa92} and
$v/\sigma^{\ast}=0.15$ \citep{frie94}. And since much, perhaps most,
of the hot gas is produced by stellar mass loss \citep{math90}, the
hot gas should also rotate negligibly.  In principle, if the hot gas
flows inwards while conserving angular momentum then the resulting
spin-up could translate into dynamically important rotational
velocities. (This would also apply to any primordial gas flowing in
from large radius.)  Both theory and the observed lack of highly
flattened X-ray images of galaxies and centrally E-dominated groups
indicate that it is highly unlikely that angular momentum is conserved
in the hot gas \citep{nuls84,brig00}. At any rate, the lack of
evidence for cooling flows noted above argues against inflowing gas.
It would also be difficult to reconcile the X-ray PA twist in NGC 720
with substantial solid body rotation around a single symmetry axis.
Thus, all evidence suggests negligible rotation of the hot gas.

Finally, we shall neglect the self-gravity of the hot gas in our
calculations. We have shown previously using \rosat\ data
\citep{buot94} that the mass of the hot gas is $\la 1\%$ of the
gravitating mass over the entire region studied.

\subsection{Current Objective and Implementation}
\label{imp}

The X-ray PA twist provides immediate geometric evidence for dark
matter since the stellar isophotes do not display any substantial
twist. However, following our discussion in \S \ref{results} regarding
the removal and replacement of point sources on the derived PA values,
here we investigate whether the ellipticities and radial profile of
the X-ray image corroborate the evidence for dark matter provided by
the PA twist. We will focus our analysis of \ex\ in the region
$35\arcsec \le a \le 139\arcsec$. The lower limit is chosen to avoid
residual point source-replacement issues (\S \ref{src}). The upper
limit avoids the outer low-surface brightness regions where \ex\ and
PA diverge from their values at smaller $a$. (The origin of this
divergence is unclear. Some possibilities are discussed in \S
\ref{conc}.) Since $\radpro$ is quite insensitive to
residual source-replacement issues and small variations in \ex\ and
PA, we will use its constraints over the entire range $r \le
185\arcsec$.

The first ingredient required for the Geometric Test is \phis, the
potential computed assuming $M\propto \ls$. Using the effective radius
of $R_e=52\arcsec$ obtained from optical surface photometry of NGC 720
\citep{burs87}, we adopt a corresponding effective semi-major axis,
$a_e = R_e / \sqrt{1-\langle \eopt \rangle}$, where $\langle \eopt
\rangle=0.31$ is the intensity weighted radial average of the optical
B-band isophotal ellipticities, \eopt, computed from the tables of
\citet{pele90} and \citet{laue95} (see Figure \ref{fig.data}). We
approximate \ls\ with a spheroidal Hernquist model
\citep{hern90,hern92} with scale length, $\as=a_e/1.8153 =
34.5\arcsec$, and constant ellipticity, $\es=0.31$.

We will consider both oblate and prolate spheroids to bracket the
range of projected ellipticities of a triaxial ellipsoid. With no loss
in the robustness of the Geometric Test we shall also assume that the
symmetry axis lies in the plane of the sky.  According to the ``X-Ray
Shape Theorem'' the unknown inclination angle $i$ must be the same for
\phis\ and \jx\ if $M\propto \ls$. Since an inclined spheroid is
necessarily rounder in projection than when viewed edge-on,
deprojection assuming $i=90\degr$ will yield \phis\ and \jx\ with
rounder shapes than in actuality. This effect tends to smear out
differences in the shapes of \phis\ and \jx\ and therefore makes it
harder to falsify the $M\propto \ls$ hypothesis; i.e., the assumption
of $i=90\degr$ is conservative in this context.

After calculating \phis\ the next step is to compare the shapes of the
isopotential surfaces to the shapes of the isoemissivity surfaces of
\jx. To perform this comparison, we parameterize \jx\ with a $\beta$
model\footnote{Here the $\beta$ model is used only as a convenient and
accurate parameterization of $\radpro$. We do not attribute the
physical significance to the model (or make many of the assumptions)
that inspired its creation by \citet{beta}.} along the major axis, $z$
(i.e., the long axis in the sky plane): $\jx(0,0,z) \propto (\rc^2 +
z^2)^{-3\beta}$ where \rc\ is the ``core radius''.  If \rc\ and
$\beta$ are given, then any $\jx(x,y,z)$ can be computed by
identifying the surfaces of constant \jx\ with those of \phis; i.e.,
for any point $(x,y,z)$ we find the corresponding value $Z$ such that
$\phis(x,y,z) = \phis(0,0,Z)$.  Then we set $\jx(x,y,z)=\jx(0,0,Z)$ at
which point we add a ``$\star$'' to \jx\ (i.e., \jxs) to emphasize
that its isoemissivity surfaces have been aligned with \phis.

We obtain the appropriate values of \rc\ and $\beta$ by projecting
\jxs\ along the line of sight to obtain \sigxs. Then we compute \radpros\
for this $M\propto \ls$ model analogously to the \chandra\ data and
perform a $\chi^2$ fit to the data. Interestingly, this flattened
$\beta$ model is a slightly better fit ($\chi^2=33.9$) than the
spherical version reported in \S \ref{radpro}.  The best-fitting
parameters and $1\sigma$ errors are, $\rc=5.3\arcsec \pm 0.3\arcsec$
and $\beta=0.414\pm 0.002$. These $1\sigma$ errors are computed using
a Monte Carlo procedure. We randomize the best-fitting \radpros\
according to Gaussian statistics and fit the model as described
above. After twenty such realizations are performed we compute the
standard deviations of \rc\ and $\beta$ obtained from the 20 runs. We
take these standard deviations to be the $1\sigma$ errors.

The X-ray image ellipticities for this $M\propto \ls$ model, denoted
by \exs, are also computed from \sigxs\ analogously to the \chandra\
data. We also compute the $1\sigma$ errors on \exs\ using the 20 Monte
Carlo runs mentioned above. For the oblate $M\propto \ls$ model we
plot \exs\ and the $1\sigma$ errors in Figure \ref{fig.data} indicated
by the ``$M\propto L_{\rm stars}$ Prediction''. For most values of $a$
the $1\sigma$ errors are smaller than the size of the value marker.

Overall the values of \exs\ are smaller than observed. For $a\la
50\arcsec$, \exs\ usually lies within the error bars estimated for
\ex\ computed from the \chandra\ image. For larger values of $a$, \exs\
increasingly deviates from the observed \ex\ values; i.e., the
deviation of $M\propto \ls$ prediction from the observations becomes
pronounced outside of $\sim R_e$. The primary reason why the
$M\propto\ls$ model cannot produce the X-ray ellipticities is because
$\ls$ is too centrally concentrated. Consequently, the spherically
symmetric monopole term dominates the stellar potential and predicts
values of \ex\ that are smaller than observed, particularly for $a\ga
60\arcsec$. (Note the downturn in \exs\ at small $a$ is an artifact of
the iterative moment method used to calculate the ellipticities. The
isophotal ellipticities of the $M\propto \ls$ model continue to
increase toward the center.)

To quantify the significance of this deviation between \exs\ and \ex\
we perform a $\chi^2$ test defined in the usual way,
\begin{equation}
\chisq \equiv 
\sum_i { \left[(\exs)_i - (\ex)_i\right]^2 \over (\ssigxs)^2_i +
(\ssigx)^2_i}, \label{eqn.chi}
\end{equation}
where $i$ represents each value of $a$ included in the sum, \ssigxs\
is the standard deviation for \exs\ mentioned above, and \ssigx\ is
the standard deviation associated with \ex\ (\S \ref{method}).  But
since values (and standard deviations) of \exs\ and \ex\ for a
particular $a$ are correlated with values at adjacent inner $a$, we
cannot interpret the result with the standard $\chi^2$ null hypothesis
probability which assumes uncorrelated errors for each point.

Consequently, we construct our own \chisq\ probability function that
accounts for the special correlations associated with our analysis.
To do this we compute \chisq\ in a world where \sigxs\ is the real
X-ray image of NGC 720. That is, starting with \sigxs\ generated using
the best fitting values \rc\ and $\beta$ we compute \exs; these are
just the best-fitting values $(\exs)_i$ shown in Figure
\ref{fig.data}. Then we follow the same procedure used to compute the
uncertainties on \ex\ from the real data (\S \ref{method}); i.e., the
model counts in each \sigxs\ image pixel are randomized assuming
poission statistics, and then \exs\ is computed in the same manner as
above. Hence, for each Monte Carlo realization of \sigxs\ we obtain
ellipticities $(\exs)_i^j$, where $i$ indicates the value of $a$ and
$j$ indicates the particular realization.

We perform $10^4$ such Monte Carlo realizations. Using all of these
runs we compute the standard deviations for each $(\exs)_i$. (The
magnitudes of these $1\sigma$ errors are very similar to those
obtained for the actual data shown in Figure \ref{fig.data}.)  Now for
each realization $j$ we calculate $(\chisq)_j$ replacing $(\ex)_i$ and
$(\ssigx)_i$ in equation (\ref{eqn.chi}) with respectively
$(\exs)_i^j$ and their associated standard deviations. This set of
$10^4$ values of $(\chisq)_j$ defines the special \chisq\ probability
distribution.

\subsection{Results}
\label{res}

The \chisq\ test reveals that the oblate $M\propto \ls$ model is
inconsistent with \ex\ and \radpro\ computed from the \chandra\ image
at the 96\% confidence level. The prolate $M\propto \ls$ model is
inconsistent at the 98\% confidence level. These results correspond to
\ex\ values computed within the region $35\arcsec \le a \le
139\arcsec$ as noted at the beginning of this section. Since the
deviation between the $M\propto \ls$ model and data increases with $a$
(Figure \ref{fig.data}), the significance of the discrepancy increases
if we exclude data at smaller $a$; e.g., if we consider instead \ex\
values computed within the region $66\arcsec \le a \le 139\arcsec$,
the oblate (prolate) $M\propto \ls$ model is inconsistent with the
\chandra\ data at the 98\% (99\%) confidence level. (If all \ex\
values computed within $a=185\arcsec$ are used then the magnitude of
the discrepancy is $>99.99\%$ for both oblate and prolate models.)

Thus, using only the ellipticities to quantify a discrepancy, the
Geometric Test demonstrates that the $M\propto \ls$ hypothesis is
unsatisfactory and that dark matter is required -- independent of the
temperature profile of the hot gas. This corroborates the evidence for
dark matter provided by the PA twist. As discussed above, such
``Geometric'' evidence for dark matter cannot be explained by general
modified gravity theories such as MOND.

\section{Temperature Profile}
\label{temp}

Unlike the Geometric Test, to construct detailed models of the
gravitating matter distribution we must have some knowledge of the
temperature profile of the gas. We extracted the ACIS spectra in three
circular annuli containing $\approx 3000$ background-subtracted counts
over 0.3-3 keV: $0\arcsec-30\arcsec$, $30\arcsec-70\arcsec$, and
$70\arcsec-140\arcsec$. We fit these spectra with an \apec\ thermal
plasma\footnote{http://hea-www.harvard.edu/APEC/} modified by Galactic
absorption ($\nhgal=1.55\times 10^{20}$~\cmsq). The spectral fitting
was performed with \xspec\ v11.1.0v \citep{xspec}.

Attempts to fit this model to the 0.3-3 keV data were unsuccessful for
two reasons. First, below $\sim 0.5$ keV the data lie well below the
model and can only be reproduced if absorption column densities of
$\nhgal\approx 10\times 10^{20}$~\cmsq are fitted. This effect is
undoubtedly related to existing problems in the low-energy
calibration mentioned on the \chandra\ web site. We suspect that
because the diffuse gas of NGC 720 has a relatively low temperature ($\sim
0.6$~keV) this calibration problem is more serious for NGC 720 than
hotter systems. And it is more serious for the diffuse gas than for
the harder discrete sources \citep[see][]{tesla}.

If we exclude data below 0.7 keV then much better fits can be obtained
for Galactic absorption. However, there is also excess emission above
the model for energies $\ga 2$~keV. Since this excess is most
pronounced in the outer annulus it is very likely associated with the
background normalization. (Recall that the background varied
throughout the observation -- \S \ref{obs}.) To largely avoid this
excess we then excluded data above 2 keV.

Fitting the models over 0.7-2 keV yields temperatures of $\sim
0.6$~keV in each annulus. Since, however, we have restricted the
bandwidth to such a narrow region (which also does not enclose the
energy corresponding to the best-fitting temperature) the constraints
on the spectral model are relatively weak. In particular, the
best-fitting metallicity is $\sim 0.1\solar$ but it is not well
constrained and depends sensitively on the chosen upper limit of the
bandpass. The temperatures have best-fitting values of 0.62, 0.58, and
0.62 keV respectively in the three annuli; each has $1\sigma$ errors
of $\approx 0.1$~keV. These values are fully consistent with the
\rosat\ and \asca\ values obtained in these regions
\citep{buot94,buot97a,buot98c} 

Hence, the \chandra\ data confirm that the hot gas in NGC 720 is
consistent with being isothermal. Given the extenuating issues
mentioned above, this is the only robust conclusion we believe can be
drawn from the spectrum of the diffuse gas at this time. 

\section{Ellipticity of the Dark Matter Halo}
\label{dm}

The Geometric Test established the need for a dark matter halo that is
distributed differently from the stars. Now we wish to find the range
of halo ellipticities that are consistent with the \chandra\
data. Since we are concerned primarily with the flattening of the halo
we shall follow the approach adopted in \S \ref{geomtest} and consider
both oblate and prolate spheroids to bracket the range of projected
ellipticities of a triaxial ellipsoid. We defer consideration of the
more computationally expensive triaxial models to a future study, but
in \S \ref{conc} we discuss briefly the suitability of a previously
published triaxial model for NGC 720.

\subsection{Preliminaries}
\label{prelim}

To constrain the halo ellipticities we follow the procedure outlined
in our previous related studies
\citep[e.g.,][]{buot94,buot96a,buot98a,buot98d} based on the
pioneering approach of \citet{bs}. We solve the equation of
hydrostatic equilibrium for the gas density assuming a single-phase,
isothermal, ideal gas,
\begin{equation}
\tilde{\rho}_{\rm g}=\exp \left[\Gamma\left(\tilde{\Phi}-1\right)\right], 
\label{eqn.gas} 
\end{equation}
where $\Gamma=-\mu m_{\rm p}\Phi(0)/k_{\rm B} T$ and where, e.g.,
$\tilde{\rho}_{\rm g}=\rho_{\rm g}(\vec x)/\rho_{\rm g}(0)$. The
assumption of an isothermal gas is an excellent approximation for
obtaining constraints on the halo ellipticity of NGC 720\footnote{In
principle the method used to test the $M\propto\ls$ model in \S
\ref{geomtest} can be applied to any mass model to constrain its
ellipticity and scale length. This approach has the advantages of not
requiring one to specify (1) the temperature profile and (2) how the
gas density $\rho_{\rm g}$ is disentangled from \sigx. The primary
disadvantage is that the method based on the Geometric Test does not,
for a general mass model, constrain the mass scale length as well as
the approach adopted in this section.}. It has been shown previously
by several studies \citep{sb,frg,buot94,buot95a} that constraints on
the ellipticity of the gravitating matter assuming an isothermal gas
are very accurate even if there is a sizable temperature
gradient. (This is a direct consequence of the X-ray Shape Theorem
noted in \S \ref{geomtest}.) In particular, for NGC 720 the \chandra\
spectral constraints (\S \ref{temp}) indicate a temperature profile
that cannot deviate substantially from isothermal. And as we have
shown previously using polytropic models of the \rosat\ PSPC data,
such small deviations from isothermality have a negligible impact on
the derived halo ellipticities \citep{buot94}.

A model X-ray image is created by integrating $\jx=\tilde{\rho}_{\rm
g}^2$ along the line of sight. (We ignore any spatial variations in
the plasma emissivity convolved with the spectral response of the
ACIS-S3 detector. Since the spectral constraints indicate the gas is
near isothermal such variations in the convolved plasma emissivity are
unimportant for our calculations.)  As in \S \ref{geomtest} we will
assume the galaxy is viewed edge-on; i.e., we do not attempt to uncover
the true inclination in this analysis. Because the outermost optical
isophotes are quite flattened (ellipticity $\sim 0.45$) it is unlikely
that the symmetry axis is significantly inclined to the line of
sight. For such small inclination angles the halo ellipticity deduced
from the X-ray analysis assuming $i=90\degr$ is a very good
representation of the true value \citep[e.g., ][]{sb,buot95a}.

For studying the ellipticity of the dark matter halo we focus our
attention on models where the mass is stratified on concentric,
similar spheroids (``spheroidal mass distributions''). We focus on the
following density profiles having asymptotic slopes spanning
the interesting range -2 to -4: (1) $\rho\propto (a_s^2 + a^2)^{-1}$, (2)
NFW model, $\rho\propto a^{-1}(a_s + a)^{-2}$ \citep{nfw}, and (3)
Hernquist model, $\rho\propto a^{-1}(a_s + a)^{-3}$
\citep{hern90,hern92}. Here $a$ is the semi-major axis and $a_s$ is a
scale length. For comparison we shall also summarize results for mass
distributions corresponding to potentials that are themselves
stratified on concentric, similar spheroids (``spheroidal
potentials''). Such potentials are generated by mass distributions
whose ellipticity varies as a function of $a$; e.g., the logarithmic
potential, $\Phi\propto \ln(a_s^2 + a^2)$, discussed by \citet{bt}. 

Given a particular mass model we compute (1) its potential, (2)
$\rho_{\rm g}$, and then (3) the model X-ray image as described
above. Next we compute the ellipticities and radial profile of the
model image just as done for the actual \chandra\ data. We obtain the
optimum model parameters by performing a $\chi^2$ fit of the model
ellipticities and radial profile to the data. The free parameters for
the models are $\epsilon$, $a_s$, $\Gamma$, and the normalization of
the radial profile. Here $\epsilon$ is the ellipticity of the mass for
the spheroidal mass distributions or the ellipticity of the potential
for the spheroidal potentials.

We calculate errors on the model parameters using a Monte Carlo
procedure similar to that done for the Geometric Test. For a given
mass model (e.g., NFW) the \radpro\ and \ex\ values of the
best-fitting model serve as the reference values.  Then we randomize
each of these reference values according to gaussian statistics and
perform fits to these randomized values analogously to the real
data. We calculate the standard deviation of the best-fitting model
parameters obtained from 20 Monte Carlo runs. We take these standard
deviations to be the $1\sigma$ errors. (Note that since the value of
\ex\ at a given $a$ is correlated with values at adjacent inner $a$,
this procedure slightly overestimates the errors on \ex. This
primarily translates to a slight overestimate of the $1\sigma$ errors
for $\epsilon$.)

\subsection{Results}
\label{re}

\begin{figure*}[t]
\parbox{0.49\textwidth}{
\centerline{\psfig{figure=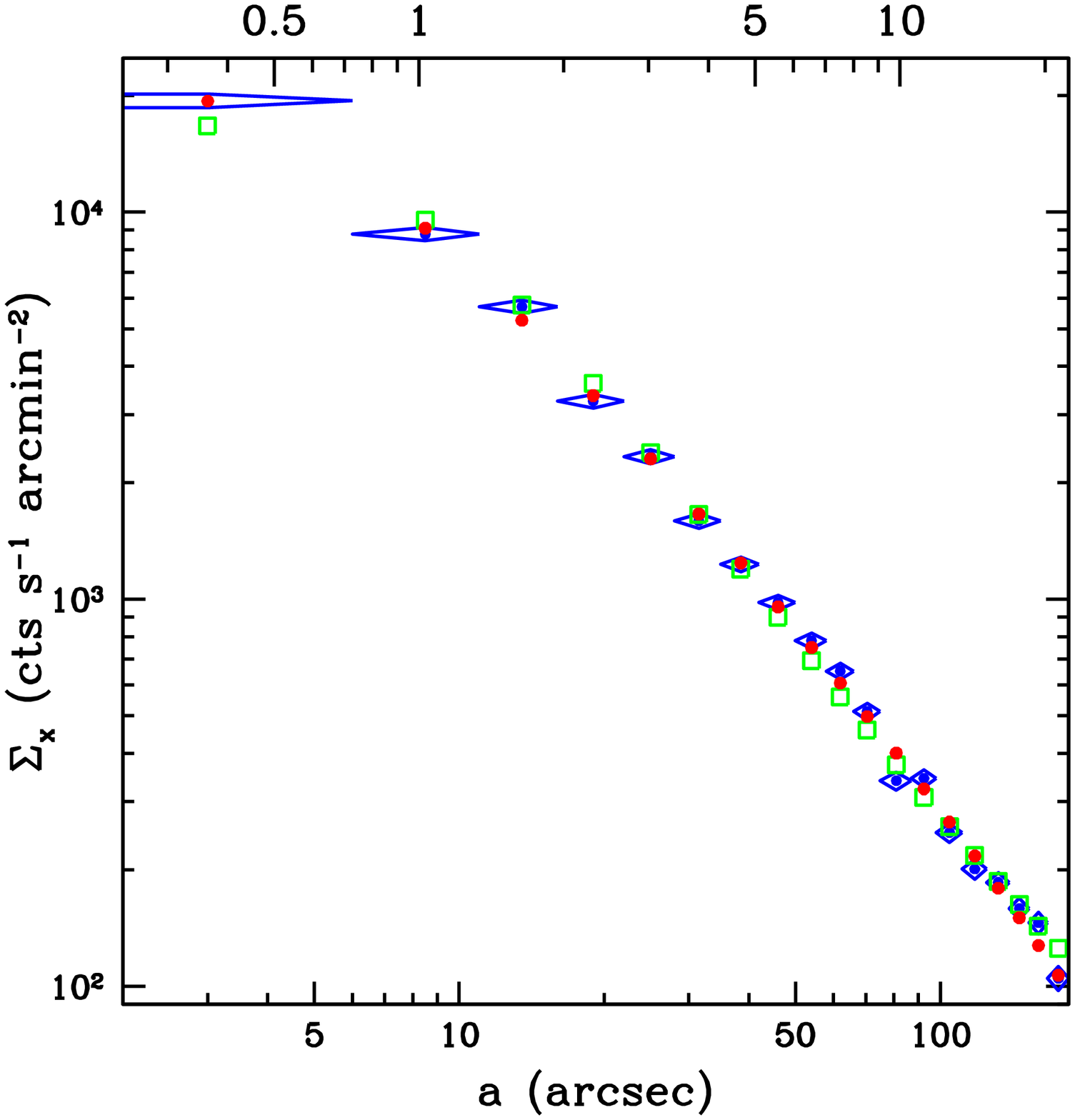,height=0.27\textheight}}
}
\parbox{0.49\textwidth}{
\centerline{\psfig{figure=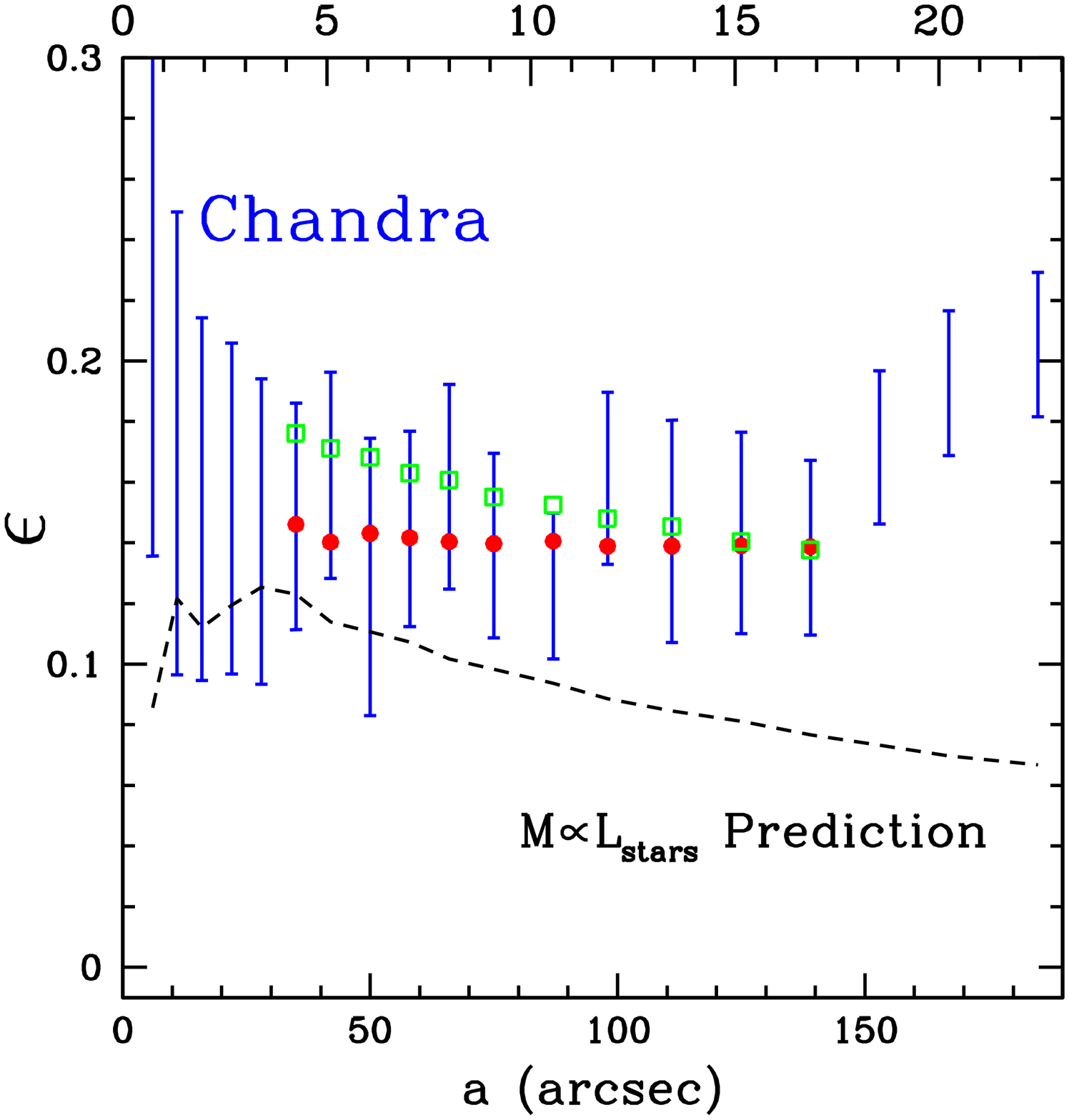,height=0.27\textheight}}
}
\caption{\label{fig.mod} ({\sl Left panel}) Radial profile denoted by
diamonds (blue). Also shown are the best-fitting
radial profiles generated by (oblate) DM halos corresponding to a
$\rho\sim a^{-2}$ profile (filled circles -- red) and an NFW profile
(open squares -- green). ({\sl Right panel}) X-ray ellipticities
predicted by these DM models. The error bars (blue) are the values of
\ex\ measured from the source-free \chandra\ data, and the dashed line is the
prediction if mass follows the stars as in Figure \ref{fig.data}. We
express $a$ in kpc on the top axis.}
\end{figure*}

\begin{table*}
\begin{center}
\caption{Ellipticity of the Dark Matter Halo}
\label{tab.mod}
\begin{tabular}{l|cccc|cccc}  \tableline\tableline\\[-7pt]
& \multicolumn{4}{c}{Oblate} & \multicolumn{4}{c}{Prolate}\\
& & $a_s$ & & & & $a_s$\\
Model & $\epsilon$ & (arcsec) & $\Gamma$ & $\chi^2$ & $\epsilon$ &
(arcsec) & $\Gamma$ & $\chi^2$\\ \tableline\\[-7pt] 
$\rho\sim a^{-2}$ & $0.37\pm 0.03$ & $1.9\pm 0.2$ & $7.95\pm 0.05$ &
32.6 & $0.36\pm 0.02$ & $2.1\pm 0.2$ & $7.82\pm 0.05$ & 32.4\\
NFW & $0.41\pm 0.02$ & $26.6\pm 0.7$ & $5.95\pm 0.03$ & 69.8 &
$0.39\pm 0.02$ & $29.3\pm 0.6$ & $5.95\pm 0.03$ & 72.5\\
Hernquist & $0.38\pm 0.03$ & $58.0\pm 1.2$ & $5.46\pm 0.03$ & 99.3 &
$0.36\pm 0.02$ & $63.4\pm 1.3$ & $5.49\pm 0.03$ & 102.2\\
\tableline \\[-1.5cm]
\end{tabular}
\tablecomments{Best-fitting values and $1\sigma$ errors obtained for
selected dark matter halo models. The mass distributions are perfect
spheroids with ellipticity denoted by $\epsilon$; see \S \ref{prelim}
for definitions of $a_s$ and $\Gamma$. There are 26 degrees of freedom
associated with each fit.}
\end{center}
\end{table*}

In Table \ref{tab.mod} we list best-fitting parameters and $\chi^2$
values for the ``spheroidal mass distribution'' dark matter
models. Technically, such single-component mass models represent the
total gravitating matter (stars, gas, dark matter) in the
system. However, as we discuss below, the dark matter sufficiently
dominates the potential so that the models in Table \ref{tab.mod}
effectively represent the dark matter. Only \ex\ values for
$35\arcsec\le a\le 139\arcsec$ are included in the fits for the same
reasons as stated for the Geometric Test (see \S \ref{imp}).

The models give consistent best-fitting values of $\epsilon\approx
0.35-0.40$ with $1\sigma$ errors of magnitude 5\%-8\%. The lowest
ellipticity estimate is provided by the oblate $\rho\sim a^{-2}$
model for which the $3\sigma$ error range is $0.27\le\epsilon\le
0.45$. The oblate NFW model provides the highest ellipticity estimate
with a $3\sigma$ error range $0.35\le\epsilon\le 0.47$. Overall,
though, there is very little difference in the fitted parameters for
oblate versus prolate models.

Previously, in \citet{buot97a} we presented our final analysis of the
\rosat\ PSPC data of NGC 720 and inferred $0.44\le\epsilon\le 0.68$
(90\% conf.) for the $\rho\sim a^{-2}$ model. These ellipticities are
systematically higher because the point sources discovered (and
removed) by \chandra\ were unresolved and therefore contaminated the
PSPC analysis. However, in that same paper we also estimated the
effect of such unresolved discrete sources on $\epsilon$ by assuming
the discrete sources would have the same spatial distribution as the
optical light. We normalized the X-ray emission of such a model using
the hard ($T\ga 5$~keV) bremsstrahlung component discovered by \asca.
For our maximal discrete model we obtained $0.29\le\epsilon\le 0.60$
(90\% conf.) which is fully consistent with the \chandra\ values.

Recall that the average B-band isophotal ellipticity is
$\langle\epsilon_{\rm opt}\rangle=0.31$ which is near the $3\sigma$
lower limit for the $\rho\sim a^{-2}$ models. The maximum optical
ellipticity of $\approx 0.45$ (see Figure \ref{fig.data}) is near the
$3\sigma$ upper limit for all the models in Table
\ref{tab.mod}. Hence, the ellipticity of the dark matter inferred from
the \chandra\ image is not very different from that of the optical
light.

The $\rho\sim a^{-2}$ model provides the best $\chi^2$ fits, and the
quality of the fit is progressively worse for models with steeper
density profiles; i.e., the Hernquist model fits even worse than the
NFW model. The NFW and Hernquist models produce lower quality fits in
part because they predict radially decreasing \ex\ profiles in
contrast to the essentially constant \ex\ profile generated by the
$\rho\sim a^{-2}$ model which agrees better with the
\chandra\ data (Figure \ref{fig.mod}). More importantly, the NFW and
Hernquist models have difficulty fitting the inner bins of the radial
profile. (The Hernquist model deviations from the data are similar to,
but larger than, those of the NFW model shown in Figure
\ref{fig.mod}.)  If the 3 inner bins are excluded the NFW model fits
as well, yields the same $\epsilon$ values, and predicts essentially
the same constant \ex\ profile, as the $\rho\sim a^{-2}$ model. This
behavior is easily understood because the best-fitting scale length,
$a_s=42\arcsec$, is also much larger than before which in effect
flattens out the NFW density profile so that approximately $\rho\sim
a^{-2}$ over most of the image.

Let us now consider the effect of the mass associated with the visible
stars on the inferred ellipticity of the dark matter. At the very
least we know that the values of $\epsilon$ in Table \ref{tab.mod}
represent lower limits for the dark matter. This is because mass
distributed like the stars predicts \ex\ values smaller than observed
(Figure \ref{fig.data}). So if the stars make a sizable contribution
to the shape of the gravitational potential, then the dark matter will
have to be flatter than inferred in Table \ref{tab.mod} to compensate
for the rounder stellar potential. (Note the mass of hot gas is much
less than the gravitating mass over the region being considered -- see
\citealt{buot94}).

We also know from the Geometric Test, and from the results for the
Hernquist models in Table \ref{tab.mod}, that a stellar mass model is
a very poor fit to the \chandra\ data. Recall from \S \ref{imp} that
the stellar mass is represented by a Hernquist model with
$a_s=34.5\arcsec$ and $\epsilon=0.31$. In Table \ref{tab.mod} we see
that even the best Hernquist models provide the worst fits, and since
the stellar values of $\epsilon$ and (especially) $a_s$ are many
standard deviations away from the best-fitting values in the table, it
is clear that the stellar model is a far worse fit to the data than
the best $\rho\sim a^{-2}$ model. It follows that the stars cannot
make a substantial contribution to the potential defined by the models
in Table \ref{tab.mod} as we found previously when modeling the
\rosat\ PSPC data \citep{buot94}. Thus, the models in Table
\ref{tab.mod} should be associated with the dark matter.

It is interesting to consider this conclusion in the context of the
mass-to-light ratio, $M/\lb$. For a distance of $25h_{70}^{-1}$~Mpc
the B-band luminosity of NGC 720 is $\lb=3.3h_{70}^{-2}\times
10^{10}\lsun$ \citep{donn90}. Assuming a gas temperature of
$T=0.6$~keV (\S \ref{temp}), the best-fitting oblate $\rho\sim a^{-2}$
dark halo model gives $M/\lb$ values in units of
$h_{70}\msun/\lsun$ of 6.0 at $r=R_e$, 12.3 at $r=2R_e$, and 18.7
at $r=3R_e$; here the masses are spherically averaged and
$R_e=52\arcsec=6.2$~kpc is the optical effective radius used to define
the stellar model in \S \ref{imp}. For comparison, the best-fitting
oblate NFW model gives $M/\lb$ values in solar units of 6.8 at
$r=R_e$, 12.7 at $r=2R_e$, and 17.0 at $r=3R_e$.

For $r=R_e$ these values are comparable to the value of $M/\lb\approx
7\msun/\lsun$ determined from stellar dynamical studies within $1R_e$
\citep[e.g.,][]{mare91}. The agreement of the X-ray and optical values
lends further support to the hydrostatic equilibrium assumption. Since
the X-ray ellipticities and radial profile (as mentioned above)
indicate that the dark matter dominates the mass over most of the
region we have studied, the agreement between the X-ray and stellar
dynamical masses of NGC 720 also implies that the stellar dynamical
mass estimate of NGC 720 must mostly reflect the dark matter and not
simply the stellar mass.

Finally, the ``spheroidal potential'' models give results fully
consistent with the spheroidal mass distributions. For example, for
the model, $\Phi\propto \ln(a_s^2 + a^2)$, we obtain (for the oblate
case) $a_s=5.2\arcsec\pm 0.3\arcsec$, $\Gamma=5.33\pm 0.05$, and an
ellipticity of the potential of $0.15\pm 0.01$. Therefore, the
best-fitting ellipticity of the dark matter is 0.20 for $a\ll a_s$ and
0.43 for $a\gg a_s$; i.e., $\epsilon\approx 0.35-0.40$ over most of
the image.

\section{Conclusions}
\label{conc}

We have analyzed a new \chandra\ ACIS-S observation of the elliptical
galaxy NGC 720 to verify the existence of a dark matter halo and to
measure the flattening of the dark halo. The ACIS image reveals over
60 point sources embedded in the diffuse emission from the hot
gas. Most of these sources were not detected with previous X-ray
observations, and analysis of their properties is presented in a
separate paper \citep{tesla}. We found it difficult to accurately
replace point sources located near the center of the galaxy with local
diffuse emission which hindered reliable computation of
\ex\ and PA for $a\la 30\arcsec$. This was not a problem for larger
$a$ where the standard \ciao\ software tool \dmfilth\ was able to
accurately replace point sources with local diffuse emission.

For $30\arcsec\la a\la 150\arcsec$ the X-ray ellipticity is consistent
with a constant value, $\ex\approx 0.15$, which is systematically less
than the values 0.2-0.3 obtained from previous \rosat\ PSPC and HRI
observations. This discrepancy can be attributed directly to the
unresolved point sources in the \rosat\ observations. The magnitude of
the PA twist discovered by \rosat\ is confirmed by the \chandra\ data:
$\pa \approx 110\degr$ for $70\arcsec\la a\la 150\arcsec$ and
$\pa\approx 140\degr$ (similar to optical value) for $a\la
30\arcsec$. However, the twist between $a=30\arcsec-60\arcsec$ is
gradual in the source-free \chandra\ data in contrast to the sharp
jump at $a\approx 60\arcsec$ observed by the \rosat\ HRI. The behavior
seen by the HRI is also attributed directly to unresolved point
sources.

\subsection{Evidence for a Triaxial Dark Matter Halo}
\label{triaxial}

The X-ray PA twist for $a\la 150\arcsec$ provides immediate evidence
for dark matter because the optical isophotes display no evidence for
such a twist. (This is a consequence of the ``X-ray Shape Theorem''
which assumes that the hot gas is single-phase -- but not necessarily
isothermal -- and in hydrostatic equilibrium -- see \S
\ref{geomtest}.) The revised PA twist also provides strong evidence
that the dark matter is triaxial. Previously, \citet{roma98}
constructed triaxial models of the \rosat\ PSPC and HRI data of NGC
720. Their best model produced a PA twist of $\sim 20\degr$ such
that $\pa\approx 140\degr$ at the center and $\pa\approx
120\degr$ for $a\ga 60\arcsec$. In Romanowsky \& Kochanek's model
the PA falls immediately, though gradually, from $a=0$ until it levels
off for $a\ga 60\arcsec$. This gradual twist disagreed strongly with
the sharp jump displayed by HRI twist near $a\approx 60\arcsec$ which
the \chandra\ observation has shown to be the result of unresolved
point sources in the HRI data.

Visual comparison of the \chandra\ PA twist (Figure \ref{fig.data}) to
Romanowsky \& Kochanek's best triaxial model (see their figure 7)
reveals that they agree quite well for $a\la 150\arcsec$. In fact, we
expect that even better agreement would be achieved with the \chandra\
data because Romanowsky \& Kochanek had to try to match the larger
\rosat\ ellipticities which inhibit the ability to produce a large
PA twist -- a point we originally made in \citet{buot96d} for NGC
720. Thus, the more gradual X-ray PA twist revealed by \chandra\
provides strong evidence for a triaxial dark matter halo and lends
important support to the assumption of hydrostatic equilibrium for
$a\la 150\arcsec$.
 
\subsection{Ellipticity of the Dark Matter}
\label{ellipdm}

To constrain the ellipticity of the dark matter halo we considered
both oblate and prolate spheroidal mass models to bracket the full
range of (projected) ellipticities of a triaxial ellipsoid. Spheroidal
models have the important advantage over triaxial models of being
less expensive to compute.

First, to complement the evidence for dark matter provided by the PA
twist, we performed a ``Geometric Test for Dark Matter'' to examine
whether mass distributed like the stars ($M\propto \ls$) can match the
observed X-ray ellipticities (irrespective of the PA twist). Focusing
on \ex\ measured for semi-major axes $35\arcsec\la a\la 139\arcsec$ we
find that the $M\propto \ls$ hypothesis is inconsistent with the
\chandra\ ellipticities at the 96\% confidence level for oblate models
and at the 98\% confidence level for prolate models. {\it Thus, both
the PA twist and the ellipticities of the \chandra\ image imply the
existence of dark matter independent of the temperature profile of the
gas. This evidence for dark matter cannot be explained by general
modified gravity theories such as MOND.}

Second, for an assumed spheroidal model of the dark matter we
generated a model X-ray image to compare the model ellipticities and
radial profile to the \chandra\ data. The dark matter density model,
$\rho\propto (a_s^2+a^2)^{-1}$, provides the best fit to the data and
gives ellipticities and $1\sigma$ errors of $\epsilon=0.37\pm 0.03$
(oblate) and $\epsilon=0.36\pm 0.02$ (prolate). NFW and Hernquist
models give similar ellipticities for the dark matter. These
ellipticities are less than the values of 0.44-0.68 (90\% conf.) 
obtained from our previous \rosat\ observations which did not account
for unresolved point sources. In our previous \rosat\ and \asca\
analysis of NGC 720 which did attempt to account for unresolved point
sources assuming such sources follow the stellar light
\citep{buot97a}, we obtained $0.29\le\epsilon\le 0.60$ (90\% conf.)
which is fully consistent with the (source-free) \chandra\ values.  It
follows that unresolved sources remaining in the \chandra\ image
should have a negligible impact on the dark matter ellipticities
determined in this paper. (This conclusion can be checked directly
with a deeper \chandra\ observation so that point sources can be
detected to a lower flux limit.)

The ellipticity of the dark matter inferred from the \chandra\ data is
not very different from the optical light, $\langle\epsilon_{\rm
opt}\rangle=0.31$ and $\epsilon_{\rm opt}^{\rm max}\approx 0.45$. But
this does not contradict the results from the Geometric Test.  The
primary reason why the $M\propto\ls$ model cannot produce the X-ray
ellipticities is because $\ls$ is too centrally
concentrated. Consequently, the spherically symmetric monopole term
dominates the stellar potential and predicts values of \ex\ that are
smaller than observed, particularly for $a\ga 60\arcsec$.

The NFW and Hernquist mass models do not fit nearly as well as the
$\rho\sim a^{-2}$ model, this discrepancy primarily arising from the
radial profile in the inner regions.  If the three inner bins are
excluded from the fits (i.e., $a\le 16\arcsec$) then the NFW model
fits as well as the $\rho\sim a^{-2}$ model. But the resulting larger
scale-length of the NFW model implies that the NFW density behaves
approximately as $\rho\sim a^{-2}$ over the $a$ range fitted. So
unless the mass actually consists of two components, i.e.,
``something'' plus NFW, the $\rho\sim a^{-2}$ model is clearly
favored. This ``something'' would have to be more centrally
concentrated than the optical light (which follows an $R^{1/4}$ law
with $R_e=52\arcsec$).

\subsection{Strange Behavior at Large Radius}
\label{strange}

At large radius ($150\arcsec\la a\la 180\arcsec$) \ex\ increases
steadily to a value of $\approx 0.20$ while the position angle
decreases to a value of $\approx 80\degr$. These changes are highly
statistically significant, although (1) the fact that this region is
near the edge of the CCD, and (2) the PAs measured by \rosat\ and
\chandra\ in this region do not agree, suggest that an independent
observation with the wide-field CCDs on \xmm\ should be used to verify
the results in this region.

If the \chandra\ values for \ex\ and PA are correct in this region,
then a physical interpretation is necessary. We have argued that there
are several good reasons to expect that the hot gas is in hydrostatic
equilibrium (see below). However, it is possible that the hydrostatic
assumption is valid only for $a\la 150\arcsec$ where the PA twist is
consistent with the triaxial dark matter halo of \citet{roma98}; i.e.,
the divergence in PA for larger $a$ could signal the departure from
hydrostatic equilibrium.

If the hydrostatic condition does hold for $a\ga 150\arcsec$ then
there must be another, larger scale, component in the dark matter.
The possibility of large-scale dark matter in NGC 720 has been
suggested from analysis of its dwarf satellites by \citet{dres86}. Why
such a separate dark component decides to assert itself at $a\approx
150\arcsec$ is unclear.

\subsection{Hydrostatic Equilibrium vs.\ Merger}
\label{he}

Since the evidence for dark matter in NGC 720 provided by the
\chandra\ data rests primarily on the assumption that the hot gas is
in hydrostatic equilibrium, here we summarize the arguments in favor
of equilibrium: (1) external influences on the diffuse gas should be
neglible because NGC 720 is very isolated (\S \ref{idea}), (2) other
than the PA twisting, there are no notable asymmetries in the X-ray
isophotes -- in particular no substantial centroid shifts (\S
\ref{idea}), (3) an N-body / hydrodynamical simulation suggests for a
system without obvious subclustering the X-ray shape analysis should
be valid (\S \ref{idea}), (4) evidence for dark matter is provided
separately by \ex\ and PA (\S \ref{triaxial} and \S \ref{ellipdm}),
(5) Romanowsky \& Kochanek triaxial model explains PA twist within
$a\sim 150\arcsec$ (\S \ref{triaxial}), and (6) the gravitating mass
within $1R_e$ determined from X-rays agrees with the stellar dynamical
value (\S \ref{re}).

A final argument for equilibrium, at least in the central regions, is
provided by the ages of the stars. \citet{terl02} estimate a
relatively young age of 3-4~Gyr for the central stellar population in
NGC 720, consistent with a fairly recent merger with a gas-rich
neighbor. However, the sound crossing time of the hot gas in NGC 720
is only $\approx 0.1$~Gyr indicating that the gas has had ample time
to settle down since the starburst. (Also, in the optical NGC 720 is
classified as a ``core galaxy'' \citep{laue95} and has rounder
isophotes near the center suggesting that it has not been dynamically
disturbed in a long time according to the simulations of
\citet{ryde02}.) We conclude that for $a \la 150\arcsec$
($18.2h_{70}^{-1}$~kpc) that hydrostatic equilibrium is very likely to
hold sufficiently well for analysis of its gravitating matter
distribution.

The only evidence that may suggest a recent merger is associated with
properties at large radius. First, the PA twist at large radius ($a\ga
150\arcsec$) cannot be explained by a single triaxial ellipsoid in
hydrostatic equilibrium (\S \ref{strange})..  Second, the ``arc'' of
point sources at $a\sim 120\arcsec$ in NGC 720 discussed by
\citet{tesla} is similar to shells seen in some elliptical galaxies
which are thought to be merger remnants.

\subsection{Implications for Galaxy Formation and Cosmology}

The existence of dark matter in isolated elliptical galaxies has been
assumed by the standard Cold Dark Matter (CDM) paradigm for many
years, but primarily because of the ambiguities associated with the
analysis of stellar velocities, definitive evidence for dark matter
has been lacking. The \chandra\ observation of NGC 720 provides vital
evidence that dark matter does indeed exist in an isolated elliptical
galaxy. This geometric evidence for dark matter is apparently the
first dynamical tracer in any stellar system that has successfully
distinguished dark matter from MOND \citep[cf. ][]{sell02b}.

The evidence for flattened dark matter provided by the ellipticities of
the \chandra\ X-ray image of NGC 720 is inconsistent with the
Self-Interacting Dark Matter (SIDM) model of \citet{sper00}. The SIDM
model predicts that the dark matter halo should be spherical within a
radius of $\sim 10$~kpc in NGC 720 using the ``collisional radius''
defined by \citet{mira02}.  On the other hand, the moderate flattening
of the dark matter halo in NGC 720 is inconsistent with the highly
flattened distributions predicted by the cold molecular dark matter
model \citep{pfen94}.

The inferred ellipticity $\epsilon\approx 0.35-0.4$ of the dark matter
halo of NGC 720 is consistent with simulations of galaxy-scale halos
in a $\Lambda$CDM model, but may be larger than indicated by a
$\Lambda$WDM (i.e., ``warm'' dark matter) model \citep{bull02}. These
ellipticities also are similar to the small number of measurements
that have been made in other types of galaxies using different methods
\citep[e.g., ][]{merr02}. Finally, the evidence for triaxiality
provided by the X-ray PA twist is another important verification of
CDM models \citep[e.g., ][]{bull02}.

Although the \chandra\ observation of NGC 720 has provided important
evidence for flattened, triaxial dark matter in an elliptical galaxy,
it is essential to apply this analysis of X-ray shapes to other
galaxies (and groups and clusters) to determine whether the
conclusions drawn from NGC 720 are representative of other
systems. Given the importance of detecting and removing point sources
for the analysis of dark matter in NGC 720, it is clear that \chandra\
and its $\approx 1\arcsec$ resolution will be essential for such
studies of other galaxies.

\acknowledgements

We thank M. Krauss for helpful discussions regarding the features in
the ACIS-S exposure map, and J. Sellwood and the anonymous referee for
comments on the manuscript. TEJ was supported by an NSF fellowship.

%XXX bibtex bibliography \\
\bibliographystyle{apj}
%\bibliography{adlrefs,dabrefs}
%\bibliography{dabrefs}

\end{document}